\documentclass[%
 aip,
 amsmath,amssymb,
preprint,%
preprintnumbers,
 reprint,%
]{revtex4-1}

\usepackage[english]{babel}
\usepackage{amsmath}
\usepackage{amssymb}
\usepackage{cancel}

\usepackage{mathtools}
\usepackage[dvipsnames]{xcolor}
\definecolor{oceanboatblue}{rgb}{0.0, 0.47, 0.75}
\definecolor{orange}{rgb}{1,0.5,0}
\definecolor{goodgreen}{rgb}{0.1,0.5,0}
\definecolor{goodred}{rgb}{0.7,0,0}
\usepackage[colorlinks,urlcolor=goodgreen,citecolor=blue,linkcolor=goodred]{hyperref}
\usepackage{bm}
\usepackage{bbm}
\usepackage{units}
\newcommand{\trick}{\includegraphics[width=0.1in]{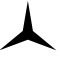}}
\newcommand{\imp}{$\boldsymbol{\psi}_{\rm R, Imp}$ }

\usepackage[T1]{fontenc}
\usepackage[utf8]{inputenc}

\newcommand{\R}[1]{{\color{goodred}#1}}

\newcommand{\tL}{\text{$t_{\rm L}$}}
\newcommand{\tR}{\text{$t_{\rm R}$}}
\newcommand{\tLR}{\text{$t_{\rm L,R}$}}

\newcommand{\ii}{\text{i}}
\newcommand{\psiR}{\text{$\bm{\psi}_{\rm R}$}}

\newcommand{\dcrit}{\text{$\delta_{\rm c}$}}

\newcommand{\comment}[1]{}
\newcommand{\Fk}[1]{\text{$ \mathcal{F}_k(#1)$}}

\newcommand{\ICSE}{\text{ICSE}}
\newcommand{\DICSE}{\text{$\delta_{\ICSE}$}}

\usepackage[normalem]{ulem}

\begin{document}

\title{Impurity-induced counter skin-effect and linear modes in the Hatano-Nelson model}
\author{Nico G. Leumer}
\affiliation{Donostia International Physics Center (DIPC), Manuel de Lardizbal 4, 20018 San Sebasti\'an, Spain}
\email{nico.leumer@dipc.org}
\author{Dario Bercioux}
\affiliation{Donostia International Physics Center (DIPC), Manuel de Lardizbal 4, 20018 San Sebasti\'an, Spain}
\affiliation{IKERBASQUE, Basque Foundation for Science, Plaza Euskadi 5, 48009 Bilbao, Spain}
\email{dario.bercioux@dipc.org}


\begin{abstract}
Non-reciprocal lattice systems are among the simplest non-Hermitian systems, exhibiting several key features absent in their Hermitian counterparts. In this study, we investigate the Hatano-Nelson model with impurity and unveil how the impurity influences the intrinsic non-Hermitian skin effect of the system. We present an exact analytical solution to the problem under open and periodic boundary conditions, irrespective of the impurity's position and strength. Numerical simulations thoroughly validate this exact solution. 
Our analysis reveals a distinctive phenomenon where a specific impurity strength, determined by the non-reciprocal hopping parameters, induces a unique skin state at the impurity site. This impurity state exhibits a skin effect that counterbalances the boundary-induced skin effect, a phenomenon we term the \emph{impurity-induced counter skin-effect}. These findings offer insights into the dynamics of non-Hermitian systems with impurities, elucidating the complex interplay between impurities and the system's non-reciprocal nature. We propose a possible implementation of this system for a non-Hermitian discrete-timequantum walk, and we demonstrate that an impurity-induced counter skin-effect also exists in multi-band models.

\end{abstract}
\maketitle

\section{Introduction}

In 1996, Hatano and Nelson~\cite{Hatano1996} demonstrated that a simple tight-binding model with non-reciprocal hopping terms can lead to unexpected localization effects of the bulk wavefunctions. This one-band model features anisotropic nearest-neighbor couplings and was originally proposed to study localization transitions in superconductors. One of the most intriguing consequences of this model is the appearance of a dramatic accumulation of eigenstates at the system boundaries when open boundary conditions (OBC) are imposed --- a phenomenon now known as the \emph{non-Hermitian skin effect} (NHSE).~\cite{yao_2018} In contrast, under periodic boundary conditions (PBC), the eigenstates remain extended throughout the bulk, a discrepancy that highlights the breakdown of the conventional bulk-boundary correspondence.~\cite{Bergholtz2021} The NHSE fundamentally arises from the non-reciprocal nature of the hopping terms, which break certain symmetries (e.g., parity $\mathcal{P}$) in the system. Typically, the effect is observed in non-Hermitian systems that exhibit a winding of the energy spectrum in the complex plane.~\cite{Okuma2020, Zhang2020, Zhang_2022c, Zhang_2022, Lin_2023} One of the main characteristics of systems displaying the NHSE is their extreme sensitivity to boundary conditions: even minor variations in boundary couplings can lead to significant changes in both the spectrum and the localization properties of the eigenstates;~\cite{Koch_2020, Lin_2023} this peculiarity is at the basis of various proposals for the detection of weak signals.~\cite{Budich_2020, McDonald_2020} The phenomenology of the NHSE has been recently extended to non-Hermitian systems with flat bands,~\cite{Martinez_Strasser_2023, Martinez_Strasser_2024, yang_2024} and fractal lattices.~\cite{Manna2023}

Recent experimental implementations of the Hatano–Nelson model have been demonstrated in various platforms, including topolectrical circuits,~\cite{Helbig_2020}ring resonator using a synthetic frequency dimension,~\cite{Wang2021} multi-terminal quantum Hall devices,~\cite{Ochkan2023} mechanical continuous systems,~\cite{Maddi_2024} and in photonic time-multiplexed resonators.~\cite{Parto2025} Moreover, the NHSE has been verified in a wide range of experimental setups, from optical waveguide arrays~\cite{weidemann_2020} to cold-atom platforms.~\cite{liang_2022}

An additional peculiarity of the non-Hermitian system is that the left and the right eigenstates are different. This leads to a fundamental question of how to evaluate physical observables in a quantum mechanical framework. The statistical interpretation of quantum mechanics fundamentally depends on the choice of metric, which is typically not unique and must be handled with care --- especially regarding left and right eigenvectors --- when deriving physical conclusions, as highlighted in studies of quasi-Hermitian Hamiltonians.~\cite{Geyer2008} A possible solution to this problem is the implementation of the so-called \emph{biorthogonal} formulation of quantum mechanics thereby setting the metric implicitly.~\cite{brody_2014, Sim_2025, Geyer2008} Here, the left and the right eigenstates are orthogonalized to each other, and physical observables are defined as an expectation value over the left and right eigenstates. This construction ensures that observable quantities (probabilities, mean values) are independent of the arbitrary overall phase between left/right eigenvector pairs. Within this formulation, the NHSE cancels since it is the opposite for left and right eigenstates. This complete treatment of non-Hermitian systems within the biorthogonal quantum mechanics differs from the Lindblad master equation formulation in which an effective non-Hermitian Hamiltonian arises as an effective description of dissipation.~\cite{Breuer_2002, Sim_2025} We emphasize that the eigenvectors of the associated Lindbladian gernally lack physical meaning. We note on passing that some recent work suggests that the NHSE is not robust to fluctuations when treated within the Lindblad master equation approach.~\cite{Ehrhardt_2024}

The localization associated with the NHSE can compete with other similar phenomena, such as the Anderson localization~\cite{Evers_2008} and the Wannier-Stark localization.~\cite{Emin_1987} In the former case, the interplay can lead to the appearance of an asymmetric Anderson localization characterized by a finite winding number and by two Lyapunov exponents,~\cite{Jiang_2019} and to chiral currents.~\cite{Sarkar_2022} In the latter case, the interplay between the NHSE and the Wannier-Stark localization leads to rich entanglement dynamics and phase transitions,~\cite{Li_2024} with the NHSE being more robust to external driving by an electric field.~\cite{Chakrabarty_2025}

It is well known that single impurities or defects in a tight-binding model can act similarly to boundaries, modifying the local density of states and potentially introducing localized modes.  In the context of the Hatano–Nelson model, this observation naturally raises the question of how impurities affect the NHSE. Under PBC, where the NHSE is absent, and the bulk states remain extended, impurities can create an \emph{internal} boundary that traps states, effectively mimicking the localization typically observed at the system edges under OBC. Several studies have explored this phenomenon, showing that even a single impurity can lead to a substantial reorganization of the eigenstate distribution, with the degree of localization depending sensitively on the impurity strength.~\cite{Li2021, Liu2021, Guo2021, Liu2020,Roccati_2021,Molignini2023,Spring2023} 
Moreover, the interplay between impurity-induced localization and the inherent non-reciprocal hopping --- which drives the NHSE under OBC, can give rise to competing localization mechanisms. Such competition may, in certain parameter regimes, result in a cancellation of the skin effect for modes localized around the impurity. In our work, we investigate this delicate balance and its implications for the spectral properties of non-Hermitian systems.

In this manuscript, we investigate the impact of a single impurity  in the Hatano–Nelson model under open boundary conditions. We show that the competition between the NHSE at the physical boundaries and at the impurity site can lead to a cancellation of the skin effect for the impurity mode. We refer to this phenomenon as the \emph{impurity-induced counter skin-effect} (ICSE). The remainder of this manuscript is organized as follows: in Sec.~\ref{sec_ii}, we present the Hatano–Nelson model with impurities and outline the solution method. In Sec.~\ref{sec_iii}, we discuss our results for impurities of various strengths for PBC and OBC, including the emergence of a linear mode and the ICSE. In the end of the result section, we present a possible implementation of the ICSE for a biased quantum walk. Finally, in Sec.~\ref{sec_iv}, we summarize our findings and offer perspectives for implementation in the framework of the non-Hermitian quantum walk.
\section{Model and Formalism}\label{sec_ii}
\begin{figure}[!t]
    \centering
    \includegraphics[width = \columnwidth]{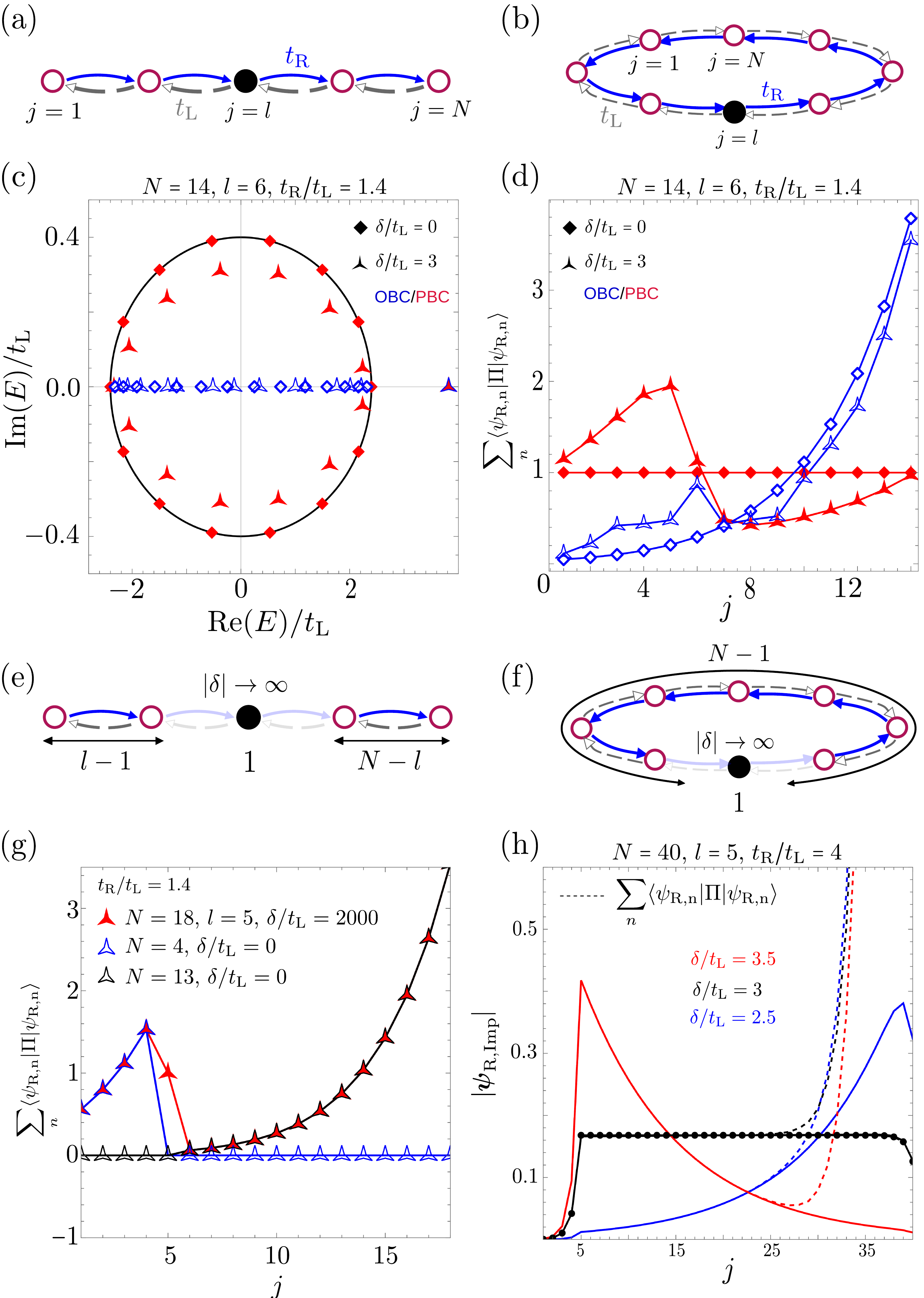}
    \vspace{-0.8cm}
    \caption{\textbf{Hatano-Nelson model with impurity: spectrum, NHSE} and impurity mode. (a) ((b)) Sketch of the system under OBC (PBC), non-reciprocal hopping $t_\text{L,R}$ and onsite impurity $\delta$ ($\bullet$).
    (c) Energy eigenvalues for $\delta/\tL =0$ ($\delta/\tL =3$) shown as $\diamond$ ($\trick$) under PBC (OBC) in red(blue). Black line: Bloch spectrum 
    from Eq.~\eqref{spectrum_PBC} for $N\rightarrow \infty$. (d) NSHE effect for right-hand eigenvectors (same color/ markers as in (c)) in agreement with Ref.~[\onlinecite{Liu2021}]. (e) ((f)) Effective chain fragmentation in the limit $\vert\delta\vert\rightarrow \infty$ for OBC (PBC).  (g) The NHSE emerges at the intermediate boundary, i.e. the impurity, due to the chain fragmentation at large $\delta$. Red: full model for $N=18$, $l=5$ and OBC. Blue/black: chain fragments with $N-l=13$, $l-1=4$ sites, $\delta =0$ and OBC. (h) Competing localization effects on $\psi_{\rm R, Imp}$ caused by NHSE and onsite impurity $\delta \neq 0$ for fixed $N=40$, $l=5$, $\tR/\tL = 4$. Strong (weak) impurities shown in red (blue) overcome (underlie) the NHSE. Intermediate values of $\delta$ may cause a compensation (black) resulting in a near flat profile of $\psi_{\rm R, Imp}$ to one site of the impurity.  Dashed lines show the NHSE respectively, and block dots mark our analytic approximation for \imp. Discussion in the main text. All data is obtained by exact diagonalization. In panels (d) and (g), the projector $\Pi$ is defined as $\Pi\equiv\Pi_n=\sum_\alpha |e_{\alpha,n}\rangle\langle e_{\alpha,n}|$, where $\Pi\equiv\Pi_n=\sum_\alpha |e_{\alpha,n}\rangle\langle e_{\alpha,n}|$ is an eigenstate of the system.}
    \label{Fig1}
\end{figure}

In the following, we will consider the Hatano-Nelson model~\cite{Hatano1996} in the presence of an onsite impurity. This one-dimensional, non-reciprocal, non-Hermitian model is characterized by hopping amplitudes different for the left (L) and right (R) hopping directions. The model Hamiltonian reads:
\begin{align}\label{eq: Hamiltonian}
    \hat{H}_l = \sum_{j=1}^{N-1} (\tL\,c_j^\dagger c_{j+1}+ \tR\,c_{j+1}^\dagger c_{j}) + \delta \,c_l^\dagger c_l
\end{align}
where $\tLR \in \mathbbm{C}\setminus\{0\}$ are the two different hopping amplitudes, and $\delta \in \mathbbm{C}$ is the impurity strength that can be placed on any lattice site $l\in\{1,\ldots,N\}$. In the previous Hamiltonian, $c_j$ ($c_j^\dag$) represents the single-particle creation (annihilation) operators for a particle at site $n$.
We present a sketch of the system in Fig.~\ref{Fig1}\R{a} and \ref{Fig1}\R{b} for OBC and PBC, respectively. The hopping terms could be fixed as $t_\text{R}=t \text{e}^{g}$ and $t_\text{L}=t \text{e}^{-g}$ with $g=1/2\ln(t_\text{R}/t_\text{L})$ and $t\in\mathbb{R}$; this choice would permit to use $t$ as a scale of energy but would not allow to have a negative product $t_\text{L}t_\text{R}$ in Eq.~\eqref{spectrum_OBC}.

Analytically, we obtained right (left) hand eigenvectors $\boldsymbol{\psi}_{\rm R}$ ($\boldsymbol{\psi}_{\rm L}$) from Eq.~\eqref{eq: Hamiltonian} solving a recursive formula of their respective entries. Since non-reciprocal hopping connects only nearest neighbors, the recursion consists of only three terms, i.e., it is of ``Fibonacci'' type\cite{Webb-1969, Hoggatt-1974, Oezvatan-2017, Leumer2023, NLthesis, Leumer21, Shin-1997, Kouachi-2008, Leumer-2020, Yueh-2005} and solutions follow similar to $\delta=0$ as superpositions of left/right moving contributions modulated by an overhaul exponential localization due to the non-Hermitian skin effect.~\cite{Liu2021, Guo2021, Edvardsson2022, Ehrhardt_2024} Also, in the language of standard 1D scattering theory,~\cite{griffiths2018} the eigenvector equation provides a continuity condition that connects left/right moving solutions before and after the impurity. Respective superposition coefficients, required to obtain the eigenvector itself, are found from the applied boundary condition either OBC or PBC. Yet, energy and associated wavevectors $k_n$ have still to be found from the (transcendental) quantization constraint --- see Eq.~\eqref{eq_two} in next the section. A full analytical Fibonacci-type solution is presented in the Supplemental Material (SM).
%
%
%
%
Further, we have verified our analytical results by considering numerical routines with infinite precision~\cite{Mathematica} since it is well known that machine precision may cause faulty data. In the SM, we show that this issue also depends on the solver/routine applied.  
\section{Results}\label{sec_iii}
We start analyzing the spectral properties of the Hatano-Nelson model by considering impurities of various strength $\delta$. We will also show the appearing of out-of-band transition, linear modes and the ICSE. For completeness, we will consider the cases both with PBC and OBC. A sketch of our model is presented in Fig.~\ref{Fig1}\R{a} and~\ref{Fig1}\R{b} under OBC, and PBC, respectively.

\paragraph{Limiting cases in $\delta$.---} The Hamiltonian comprises two important and complementary limits concerning the impurity strength that read $\delta = 0$ and $\vert \delta\vert  \rightarrow \infty$. In the former case, Eq.~\eqref{eq: Hamiltonian} reduces to the well-known Hatano-Nelson model, and under PBC, its spectrum reads 
%
%
\begin{equation}\label{spectrum_PBC}
    E(q_n) = (\tL+\tR) \cos(q_nd) + \ii (\tL-\tR) \sin(q_nd)    
\end{equation}
%
%
with $q_nd = 2n\pi/N$, $n=1,\ldots,\,N$. It winds around the origin in the complex plane whenever the hopping amplitudes differ, i.e., $\tL\neq \tR$ --- cf.~Fig.~\ref{Fig1}\R{c}, red $\diamond$. In contrast, under OBC, the spectrum reads
%
%
\begin{equation}\label{spectrum_OBC}
E(k_n) = 2\sqrt{\tL\tR} \cos(k_nd)
\end{equation}
%
%
with $k_nd = n\pi/(N+1)$, $n=1,\ldots,\,N$, and it resides on the real (imaginary axis) if $\tL\tR>0$ ($\tL\tR<0$) as shown by blue $\diamond$ in Fig.~\ref{Fig1}\R{c}. Predicted by the spectral winding for PBC, all right-hand energy eigenstates of Eq.~\eqref{eq: Hamiltonian} under OBC pile up towards the right (left) chain's end in case of $\tR/\tL>1$ ($\tR/\tL<1$) as shown in Fig.~\ref{Fig1}\R{d} --- this represent the non-Hermitian skin effect. 

We consider then, the second interesting limit $\vert \delta\vert \rightarrow \infty$. Although hopping $\tLR$ still connects all neighboring sites, the chain effectively splits into various \emph{fragments}, three in the case of OBC with lengths $l-1$, $1$, $N-l$  and two for PBC with length $1$, $N-1$, as illustrated in Fig.~\ref{Fig1}\R{e} and Fig.~\ref{Fig1}\R{f}, respectively. The reason for this is, similar to the tunneling effect in textbook quantum mechanics that the impurity acts as potential barrier $V(j)$\comment{$V(j) = \delta \,\delta_{jl}$}, i.e., that the energy difference $V(j=l)-V(j\neq l) = \delta$  does not support hybridization between regions of different onsite energies.~\cite{Roccati_2021} Indeed, the eigenvector equation from Eq.~\eqref{eq: Hamiltonian} displays explicitly an additional OBC at the impurity position for all states from the various fragments with $N-1$ sites for PBC and $l-1$, $N-l$ sites for the OBC case. The single exception is the impurity mode $\psi_{\rm R, imp}$. This state has an energy $E\simeq \delta$ and is trapped at the impurity site.

In Fig.~\ref{Fig1}\R{g}, we show the NHSE found from Eq.~\eqref{eq: Hamiltonian} for $\delta/\tL = 2000$, $l=5$, $\tR/\tL=1.4$ and $N=18$ in red. Since the impurity strength is the dominant energy scale, the illustrated data mimics the scenario of $\vert \delta\vert\rightarrow \infty$. In this regard, blue and black points correspond to impurity-free Hatano-Nelson chains for $\tR/\tL=1.4$ and respective lengths of $l-1\equiv 4$ and $N-l\equiv 13$ sites. The perfect match of blue/red (black/red) data points on the sites $j=1,2,3,4$ ($j=6,\ldots,18$) showcases that the impurity realizes OBC, i.e., the chain effectively fragments. At $j=l=5$, only the trapped impurity mode (red) contributes such that the NHSE assumes unity. 

Finally, in Fig.~\ref{Fig1}\R{h}, we show the competition between the exponential localization of the impurity and the NHSE, shown in red and blue, respectively. For a specific value of $\delta$, the two localization lengths nearly cancel on the impurity site (black) causing a nearly flat profile. This effect is the impurity-induced counter skin-effect, and we shall discuss it further down below.

\paragraph{Generic $\delta$ and quantization condition for OBC.---} Generally, the case of finite (complex) $\delta$ can be understood as an interplay between the two extreme limits discussed above and it manifests in the quantization condition of wavevectors $kd$ associated to energy $E$. In case of OBC, Eq.~\eqref{spectrum_OBC} still holds with $k_nd$ being the (complex) solutions of
\begin{align}\label{eq_two}
    \frac{\delta}{\sqrt{\tL \tR}}\,\frac{\sin(k_n d\,l)}{\sin(k_nd )} \frac{\sin[k_nd (N-l+1)]}{\sin(k_nd )} = \frac{\sin[k_nd (N+1)]}{\sin(k_nd )}.
\end{align}
Notice that at $\delta=0$, Eq.~\eqref{eq_two} reduces to the known quantization condition $\sin[k_nd (N+1)] = 0$ for the impurity-free Hatano-Nelson chain consisting of $N$ connected sites. Dividing Eq. \eqref{eq_two} by $\delta\neq0$ and assuming that $\delta \rightarrow \pm \infty$, we have $\sin(k_n d\,l)=0$ and $\sin[k_nd (N-l+1)]=0$. Since the two constraints give generally two distinct sets of solutions $k_n$ for arbitrary $N$, $l$, Eq. \eqref{eq_two} manifests explicitly the chain's fragmentation into two sub-chains of respectively $l-1$, $N-l$ sites. This limit extends to complex $\delta$ as can by seen by using the polar form $\delta = \vert \delta\vert e^{i\phi_\delta}$. In addition, the real part of solutions ${\rm Re}(k_nd) \in [-\pi/2,\pi/2)$ may be restricted to the first Brillouin zone exploiting the periodicity of the $\sin$ functions. Because the sign inversion $\delta\to -\delta$ in Eq.~\eqref{eq_two} is always counteracted by the shift $k_nd\to k_nd+\pi$, the spectrum only reverses its sign when the impurity does.

Since the wavevector quantization condition under PBC also displays the chain's fragmentation for $\vert \delta\vert \rightarrow\infty$ explicitly, Figs.~\ref{Fig1}\R{c} ,~\ref{Fig1}\R{d} can be understood as follows. In Fig.~\ref{Fig1}\R{d}, eigenvectors of Eq.~\eqref{eq: Hamiltonian} under PBC (OBC) signal the emergence of the NHSE at the impurity site --- cf. \trick. This is accompanied by a change of the associated eigenvalues shown in Fig.~\ref{Fig1}\R{c}. Although energies under PBC and $\delta\neq0$ still wind around the origin of the complex plane, the ellipses flatten and shrink. This behavior is clear when $\vert \delta\vert \rightarrow \infty$, since the fragmentation of the system is equivalent to have OBC, i.e., the spectrum becomes purely real (imaginary) whenever $\tL\tR>0$ ($\tL\tR<0$). Similarly, the spectrum of Eq.~\eqref{eq: Hamiltonian} and OBC rearranges itself on the real (imaginary) axis for finite $\delta$ in order to properly display the energies eigenvalues of the chain fragments.

\paragraph{Out-of-band transition and linear modes.---} 
We notice that under OBC and sufficiently strong but finite~$\delta$, the energy spectrum of Eq.~\eqref{eq: Hamiltonian} is not restricted to $(-2\sqrt{\tL\tR},\,2\sqrt{\tL\tR})$ as illustrated by Fig.~\ref{Fig1}\R{c}. Here, we observe the appearance of a state at energy $E\approx \delta>\tR$ for both OBC/PBC beyond the band's extremes. This corresponds to the impurity mode $\psi_{\rm R, imp}$ whose energy typically is $E_{\rm Imp}\simeq \delta$, besides a small correction due to hybridization with neighboring sites. Generally, $E_{\rm Imp}$ corresponds to a complex wavevector $kd$ such that $\psi_{\rm R, imp}$ localizes exponentially around the impurity depending on the precise value of $\delta$.

\begin{figure}[!h]
    \centering
    \includegraphics[width=\columnwidth]{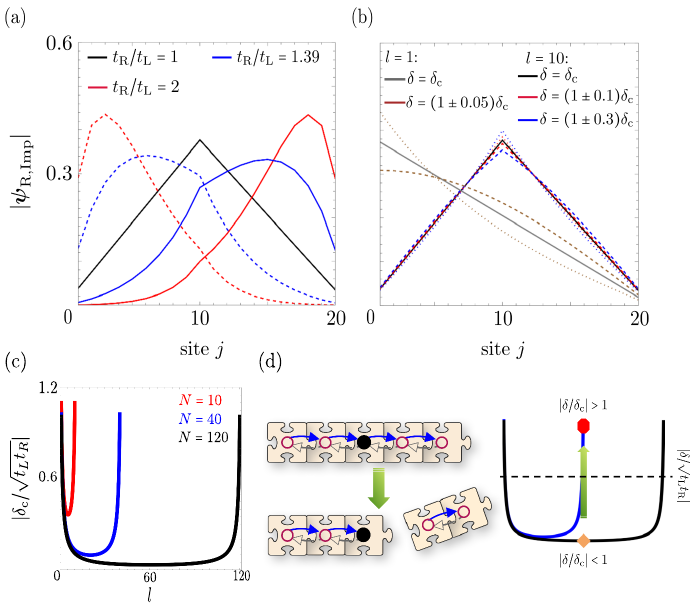}
    \caption{\textbf{Linear mode and its stability.} (a) Due to its exponential dependence, tuning the ratio $\tR/\tL$ alters $\psi_{\rm R, Imp}$'s profile (black) significantly. Data shown belongs to $l=10$, $\delta = \dcrit$ and $\tL \tR = 2$ fixed. Dashed lines belong to $\psi_{\rm L, Imp}$. (b) Modifying $\delta$ around $\dcrit$ yields minor (significant) changes for $l=N/2$ ($l=1$) comparing the blue/red (brown) lines with the black (gray) one. Dotted (dashed) curves indicate increased (reduced) $\delta$ for $\tR/\tL = 1$. In the two panels, we have used $N=20$.
    (c) $\dcrit$ as function of impurity position $l$ for different system sizes $N=10,\,40,\,120$. (d) Removing impurity-free sites may cause a transition from $\vert \delta/\dcrit >1 \vert$ to $\vert \delta/\dcrit <1 \vert$ even if the impurity strength is constant and cannot be controlled.}
    \label{Fig2}
\end{figure}
An exact solution for $\psi_{\rm R, imp}$ (in all parameter regimes), and following our analytical approach, requires merely the knowledge of the associated $kd$. However, the transcendental character of Eq.~\eqref{eq_two} prohibits an exact solution, except in certain limiting cases, such as the out-of-band transition --- here, the impurity mode energy lies beyond the band limits. Since we can continuously change parameters from $\delta = 0$ to sufficiently large values, we can identify a critical value $\dcrit$ for which the impurity modes energy equals one extreme of the band, i.e., $E_{\rm Imp} = \pm 2\sqrt{\tL\tR}$ under OBC. For this specific case, the dispersion relation demands $kd = 0,\pi$ for the associated wavevector. Therefore, the critical value
\begin{align}\label{eq_dcrit}
    \frac{\dcrit}{\sqrt{\tL \tR}} = \pm\frac{N+1}{l(N+1-l)}.
\end{align}
of $\delta$ follows from Eq.~\eqref{eq_two}, at which the oscillatory part of $\psi_{\rm R, Imp} = (\alpha_1,\ldots,\,\alpha_l, \gamma_{l+1},\ldots\gamma_{N})^\mathrm{T}$ adopts a linear shape\footnote{Linear modes do exist also in the Hermitian Su-Schrieffer-Heeger model (and the Kitaev chain for absent onsite energy) of finite size and OBC.~\cite{NLthesis} This mode appears when a former in-gap mode enters the upper/lower band by sweeping the ratio of inter/intra- band hoppings. Here, the mode may be used to define the topological phase transition under OBC and for finite chain length.}
\begin{subequations}\label{eq_linear_mode}
\begin{align}
\label{eq: lin mode alpha}
   \frac{\alpha_j}{\alpha_1} &= \left[\pm \sqrt{\frac{\tR}{\tL}}\right]^{j-1}\, j,\\
   \label{eq: lin mode gamma}
    \frac{\gamma_j}{\alpha_1} &= l
        \left[\pm \sqrt{\frac{\tR}{\tL}}\right]^{j-1}\frac{j-N-1}{l-N-1},
\end{align}
\end{subequations} 
with $j\in\{1,\ldots,N\}$ and $l$ is the impurity position. The positive sign belongs to $k_nd = 0$, whereas the negative one to $k_nd = \pi$. Notice also that $\alpha_1$ adopts the role of the normalization constant. 

In Fig.~\ref{Fig2}\R{a}, we show $\vert \psi_{\rm R, imp}\vert$ obtained numerically with infinite precision for $\delta = \dcrit$, $l=10$, $N=20$ and fixed $\tL \tR = 2$. In the Hermitian case $\tL = \tR$ (black), we witness a linear shape peaked at the impurity position $j=l=10$. In case of anisotropic hopping $\tL  \neq \tR$, the NHSE localizes $\vert \psi_{\rm R, imp}\vert$ towards the right (left) end whenever $\tR>\tL$ ($\tR<\tL$) as Eqs.~\eqref{eq_linear_mode} suggest that the exponential dominates locally over the linear term. The results for lhs eigenvectors $\vert \psi_{\rm L, imp}\vert$ (dashed) follow from those of $\vert \psi_{\rm R, imp}\vert$ by mutual exchange of $\tLR$. 

In Fig.~\ref{Fig2}\R{b}, we show $\vert \psi_{\rm R, imp}\vert$ for $\vert\delta/\delta_c\vert>1$ ($\vert\delta/\delta_c\vert<1$) in dotted (dashed) lines, while the solid ones belong to $\vert\delta/\delta_c\vert=1$. When $\delta$ exceeds the critical value $\delta_c$, the linear shape is altered into an exponential one and $\vert \psi_{\rm R, imp}\vert$ localizes towards the impurity. In gray (black), we show the case of an impurity at a edge (bulk) site $l=1$ ($l=10$). Similarly, we have an oscillatory behavior of $\vert \psi_{\rm R, imp}\vert$ for $\vert\delta/\delta_c\vert<1$. Additionally, the impurity modes' profile is sensitive to the smallest changes in $\delta$ when the impurity resides close to the chain's edges. 

The actual dependence of $\dcrit$ on the various parameters is intrinsically intriguing. In terms of the impurity's position $l$, $\dcrit$ is the product of two hyperbolas with poles at $l=0, N+1$, that means beyond the chain's terminal sites, as illustrated in Fig.~\ref{Fig2}\R{c}. Hence, the largest value $\vert\dcrit/\sqrt{\tL\tR}\vert$ is $1+\nicefrac{1}{N}$ for $l=1,N$, while placing the impurity close to the center of the chain $l=N/2$ gives $(N+1)/(\nicefrac{N^2}{4}+4)$ a vanishing small contribution for longer chains $N\rightarrow\infty$. 

This is an interesting property, in particular since the rhs of Eq.~\eqref{eq_dcrit} is formed by real quantities, i.e., the out-of-band transition survives the Hermitian limit $\delta\in \mathbbm{R}$, $\tL = t_{\rm R}^*$ of the model. That implies that the transition from $\vert\dcrit/\delta\vert<1$ to $\vert\dcrit/\delta\vert>1$ can be caused by rearranging the sites. For example, in state-of-the-art scanning tunnelling microscope experiments, atoms can be placed controlled on substrates, i.e., the chain of atoms may be broken into two pieces as sketched in Fig.~\ref{Fig2}\R{d}. This allows the control of the out-of-band transition even in case $\delta$ cannot be changed directly.

\paragraph{Impurity induced counter skin effect.---}
%
%
\begin{figure}[!t]
    \centering
    \includegraphics[width = 0.95\columnwidth]{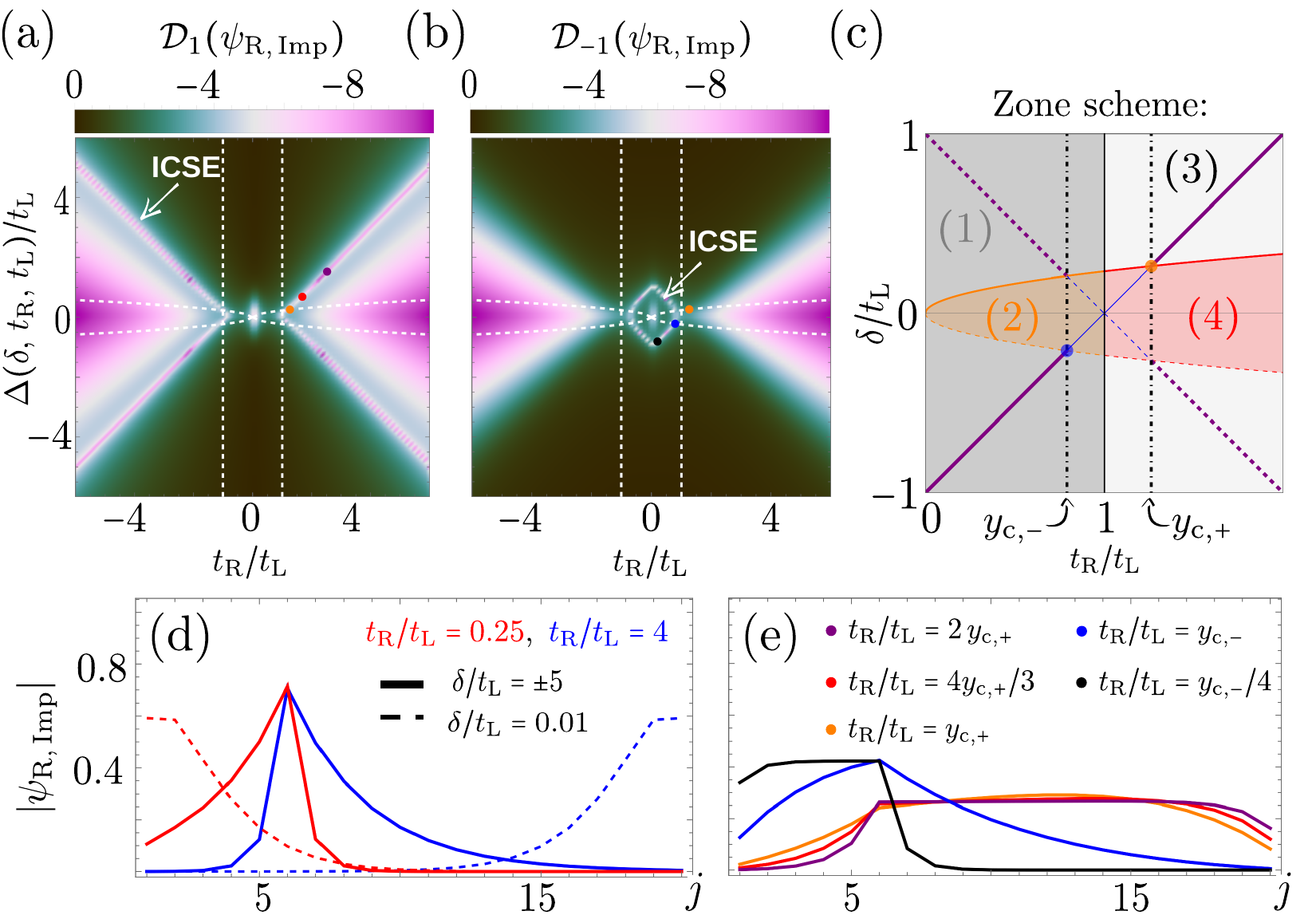}\vspace{-0.3cm}
    \caption{\textbf{ICSE phase diagram: $\mathcal{D}_{\pm 1} = \ln(10^{-5} + |D_{\pm 1}(\psi_{\rm R,\, Imp})|)$ \emph{vs} $\tR/\tL$ and $\Delta$.}  (a) ICSE (purple/white lines) confirm the linear relation $\delta = \pm (\tR -\tL)$, $\tR/\tL>0$ ($\ii \delta = \pm (\tR -\tL)$, $\tR/\tL<0$)
    for $\tR/\tL>1$ ($\tR/\tL<-1$). (b) For
    $\tR/\tL<1$, the ICSE exists as well, but detection requires $\mathcal{D}_{-1}$.
    (c) The phase diagram consists of four distinct sub-areas (1), (2), (3), (4). (d) Spatial profile of $\vert \psi_{\rm R,\, Imp}\vert$ in (1)-(4). (e) Impurity mode along the ICSE line $\delta/\tL = \tR/\tL - 1$. Discussion see main text. All data was obtained from exact diagonalization for $N=20$, $l=6$.}
    \label{Fig3}
\end{figure}
%
%
The ISCE results from a nearly perfect cancellation of exponential localizations due to the impurity and the NHSE, as shown in black in Fig.~\ref{Fig1}\R{h}. We obtain that the necessary impurity strength to observe it is
\begin{align}\label{eq: delta ICSE}
    \delta = \pm (\tR - \tL)   
\end{align}
when the impurity is not placed too close to the chain's terminal sites, cf. SM. We start by considering the case in which the impurity is placed in the center of the chain and discuss the correction caused by placing the impurity close to terminal sites later.

To verify the computational results (with infinite precision) and also to deepen our understanding of the phenomenon, we opt for an analytic approximation for $\psi_{\rm R, imp}$. Although our exact formulae for all eigenstates allow for exact results, they require the knowledge of the states' energy or the associated wavevector $kd$. Due to the complexity of Eq.~\eqref{eq_two}, we choose instead to approximate the energy directly. Thus, we return to the 
eigenvector equation from Eq.~\eqref{eq: Hamiltonian}, cf. SM, that describes the spatial change of $\psi_{\rm R, Imp} = (\alpha_1,\ldots,\,\alpha_l, \gamma_{l+1},\dots\gamma_{N})^\mathrm{T}$ behind the impurity, i.e., $\gamma_{l+1},\ldots,\,\gamma_{N}$, and apply the phenomenological constraint of a flat profile: $\gamma_{l+1} =\ldots =\gamma_N=$const. That fixes the states' total energy to be $E = \pm(\tL+\tR)$ for $\delta = \pm (\tR-\tL)$, which is very close to the actual numerical values. Using our approximated energy, Eq.~\eqref{spectrum_OBC} for $kd$ and the exact formulae for eigenstates, the approximated state is shown by the black dots in Fig.~\ref{Fig1}\R{h}; thus, confirming all computational data.

Concerning the localization strength of the impurity, we find $\kappa d = \text{arccosh}[(\tL+\tR)/(2\sqrt{\tL \tR}) ]$ using Eq.~\eqref{spectrum_OBC}, $E = \tR+\tL$ and $kd\rightarrow \ii \kappa d$ to properly account for the non-real wavevector under sufficient $\delta$ and $\tR>\tL>0$. On the basis of our assumptions, the argument of the function $\text{arccosh}$ is positive and larger than one; thus, exploiting the relation to the natural logarithm permits to obtain the length scale $1/\kappa = 2d/\ln(\tR/\tL)$. We observe that $1/\kappa$ is identical to the localization $L$ imposed by the NHSE, which can be extracted from Eqs.~\eqref{eq5} in the SM. We conclude that both effects can cancel depending on the impurities position $l$ at given $\tR/\tL$.

In the SM, we further examine the energy dependence of $\psi_{\rm R,Imp}$ when the ICSE is present. Moreover, as the ratio $\tR/\tL$ increases, the ICSE becomes progressively flatter --- resulting in a less pronounced decay, as shown by the black curve in Fig.~\ref{Fig1}\R{h}. Finally, the ICSE can also manifest to the left of the impurity; in this case, a ratio $\tR/\tL<1$ is required to shift the NHSE towards the opposite end.

To conduct a more in-depth study on the ICSE, we evaluate the following gradient of the impurity mode
\begin{align}\label{eq: definition Dx}
    D_x({\psi}_{\rm R})\coloneqq \frac{\vert\psi_{{\rm R}, l}\vert-\vert\psi_{{\rm R}, l+x} \vert}{x}
\end{align}
between sites $l$, $l+x$ as a function of $\tL$, $\tR$ $\delta$ using exact numerical results. For the largest possible range of impurity positions $l$, we show data for only $x=\pm1$. Since $D_x$ may be small due to exponential localization effects of either the NHSE or the impurity, we improve the resolution by interpreting $\mathcal{D}_{\pm}=\ln(10^{-5} + |D_{\pm1}(\psi_{\rm R,\, Imp})|)$. Also, to explore the full range of $\tLR$, we introduce 
%
%
\begin{align}
    \Delta(\delta, \tL, \tR)= \left\{\begin{matrix}
        \delta, &  \tL\tR>0\\
         \ii \delta, &\tL\tR<0
    \end{matrix}\right. .
\end{align}
%
%
This approach preserves the spectrum of Eq.~\eqref{eq: Hamiltonian} for all $\tL\tR$, since appearing complex phases in Eq.~\eqref{eq_two} cancel. Since the spectrum of Eq.~\eqref{eq: Hamiltonian} inverts its sign in case $\delta$ does, the shown data for the gradient is symmetric w.r.t. to the horizontal and vertical axes.

The results for the gradient for the impurity mode are shown in Fig.~\ref{Fig3}\R{a} for $x = +1$ and in Fig.~\ref{Fig3}\R{b} for $x = -1$ with $N=20$, $l=6$. Note that vertical dashed lines correspond to $\tL = \tR$. On the top/bottom center (dark green), the gradient registers that $\psi_{\rm R,imp}$ is exponentially localized at the impurity site $l$, while on the center left/right $\mathcal{D}_{\pm}$ (deep purple) reflects the pile up due to the NHSE. Supporting data is presented in Fig.~\ref{Fig3}\R{d}. In between both regimes and in agreement with Eq.~\eqref{eq: delta ICSE}, resides the ICSE, forming the separated and mostly purple \emph{chevron} pointing to the right and to the left in Fig.~\ref{Fig3}\R{a} and a diamond shape in Fig.~\ref{Fig3}\R{b} with detailed data in Fig.~\ref{Fig3}\R{e}.

The sketch in Fig.~\ref{Fig3}\R{c} illustrates different parameter areas inside Fig.~\ref{Fig3}\R{a} and Fig.~\ref{Fig3}\R{b}. Zones (1), (3) correspond to values of $\vert\delta/\dcrit\vert>1$, i.e., $\psi_{\rm R,imp}$ exponentially localizes at the impurity, with NHSE towards the left (right) end of the chain in (1) ((3)). Similarly in areas (2) and (4), we have $\vert\delta/\dcrit\vert<1$ and $\vert\delta/\dcrit\vert=1$ marks the boundary respectively between (1), (2) and (3), (4). In purple, we show the parameter constraint for the ICSE --- cf. Eq.~\eqref{eq: delta ICSE}. Since exponential localizations due to NHSE and impurity need to compensate for the ICSE, the latter resides actually only on the purple lines within regions (1) and (3). The transition points $y_{\rm c, \pm}$ follow from Eqs.~\eqref{eq_dcrit}, \eqref{eq: delta ICSE} and the explicit formula is given in the SM. Physically, $y_{\rm c, \pm}$ give limits for $\tR/\tL$ beyond which we expect the ICSE. 

The colored spots in Figs.~\ref{Fig3}\R{a},~\ref{Fig3}\R{b} refer to the states shown in \ref{Fig3}\R{e}. For $\tR/\tL = 2y_{\rm c, +}$ (purple), the flat profile of $\vert \psi_{\rm R, imp}\vert$ reflects clearly the ICSE which is preserved until $\tR/\tL = 4y_{\rm c, +}/3$ (red). In between $\tR/\tL = y_{\rm c, +}$ (orange) and $\tR/\tL = y_{\rm c, -}$ (blue), the ICSE is absent and, due to $\tR/\tL<1$, the NHSE now localizes $\psi_{\rm R, imp}$ towards the right end of the chain. For sufficient small $\tR/\tL = y_{\rm c, -}/4$ (black), the ICSE re-emerges.

Concerning impurities on edge sites, i.e., $l=1,N$, the ICSE does exist, but for adapted impurity strength, i.e., Eq.~\eqref{eq: delta ICSE} is no longer valid since it was derived from a bulk constraint in the absence of OBC. Instead, we find the linear relationship $\delta = \tR$ ($\delta = \tL$) for $l=1$ ($l=N$) for sufficient $\tL/\tR$. We show supporting data in the SM. In the case of impurities on the first/last few sites, the value of $\delta$ seems to transition between the respective two conditions. In the SM, we show the ICSE for a multi-band model; specifically, we consider the non-reciprocal SSH model.~\cite{Lieu_2018}
\paragraph{Application: Quantum Walk}
In this section, we demonstrate how our findings can be implemented within the framework of a discrete-time quantum walk (DTQW) under non-Hermitian dynamics.
Quantum walks generalize classical random walks by incorporating essential quantum features such as superposition and interference.\cite{Aharonov_1993,Kempe_2003} In a classical random walk, a particle moves through position space with fixed probabilities at each step. In contrast, in a quantum walk, the walker evolves coherently, exploring multiple paths simultaneously. The resulting interference of probability amplitudes—constructive or destructive—leads to markedly different behavior: in particular, the variance of the walker’s position grows quadratically with the number of steps, as opposed to the linear scaling observed in classical walks.\cite{Ambainis_2001,Nayak_2001}

The Hilbert space for the DTQW is constructed as the tensor product
%
%
\begin{equation}
\mathcal{H} = \mathcal{H}_P \otimes \mathcal{H}_C,
\end{equation}
%
%
where $\mathcal{H}_P$ denotes the position space spanned by $\{|x\rangle: x\in\mathbb{Z}\}$ and $\mathcal{H}_C\cong\mathbb{C}_2$ is the two-dimensional coin (or internal) space. Each basis element $|x\rangle\otimes|c\rangle$ (with $c=$R,L) represents the walker at postion $x$ with coin state $|c\rangle$.

A single step of the DTQW is defined by the unitary (or, in our case, non-unitary) evolution operator
%
%
\begin{equation}\label{evolution}
U = S \, \left( \mathbb{I} \otimes C \right),
\end{equation}
%
%
where $\mathbb{I}$ is the identity operator in the walker subspece, $C$ is the coin operator acting on $\mathcal{H}_C$, and
$S$ is the conditional shift operator acting on the combined space.

In the case of a Hermitian evolution of the DTQW, the coin operator can be defined as
%
%
\begin{equation}
C =  \frac{1}{\sqrt{2}}\begin{pmatrix}
1 & 1 \\
1 & -1
\end{pmatrix},
\end{equation}
%
%
in the literature know as either \emph{unbiased coin} operator\cite{Tregenna2003} or Hadamard coin operator.~\cite{Qiang2024} Later in this work, we generalize this by introducing \emph{biased} coin operators—where the amplitude for moving left or right differs—thereby laying the foundation for realizing a nonreciprocal, impurity-free analog of the Hatano–Nelson model in the DTQW framework.

The shift operator in Eq.~\eqref{evolution} is defined as
%
%
\begin{equation}
S = \sum_{x} \Big( |x+1\rangle\langle x| \otimes |R\rangle\langle R| + |x-1\rangle\langle x| \otimes |L\rangle\langle L| \Big),
\end{equation}
%
%
where $|R\rangle$ and $|L\rangle$ denote the coin states corresponding to right and left moves, respectively and after $t$ steps, the walker is
%
%
\begin{equation}
    |\psi(t)\rangle = U^t |\psi(0)\rangle.
\end{equation}
%
%
%
%
\begin{figure*}
    \centering
    \includegraphics[width=0.9\textwidth]{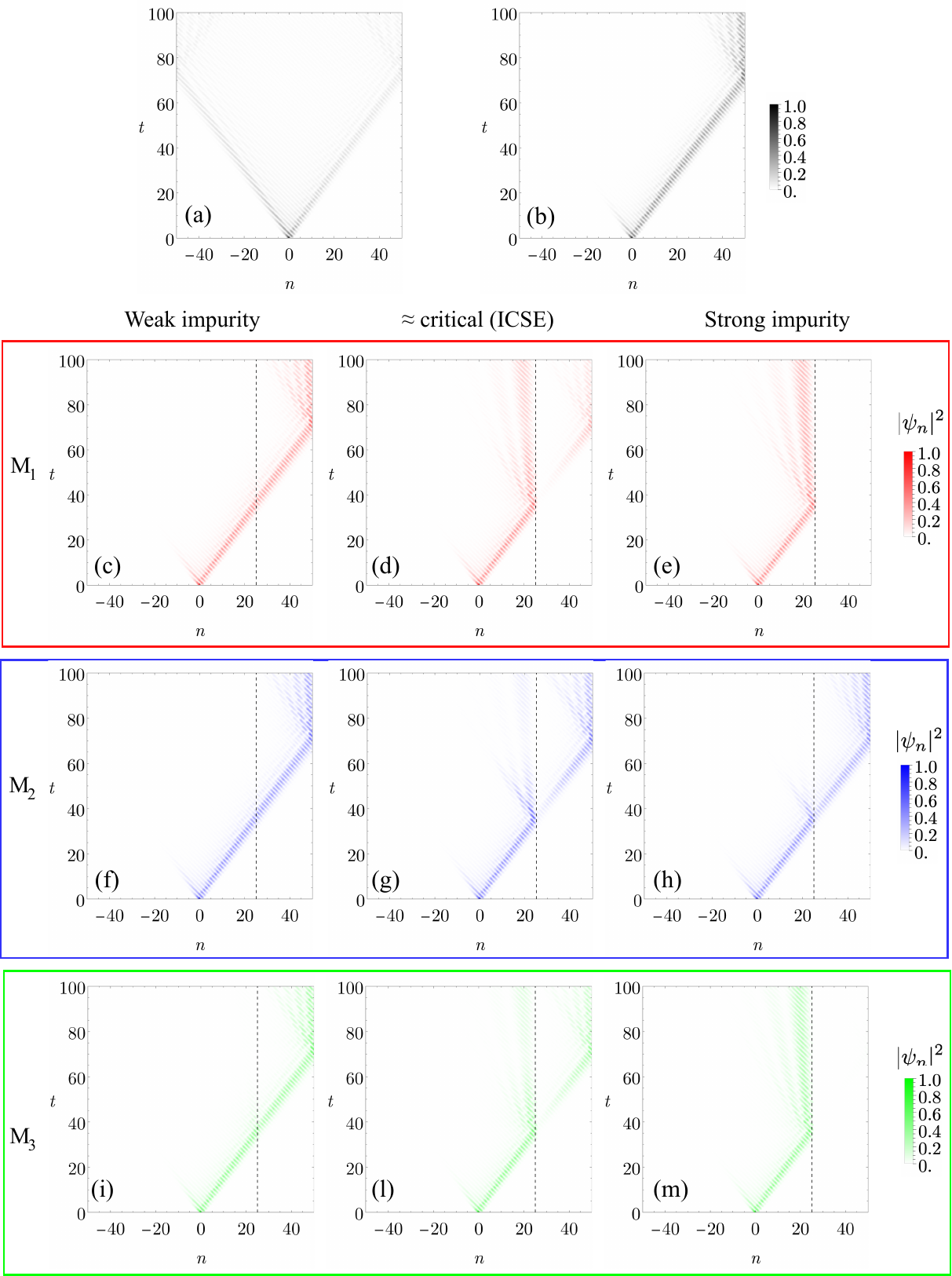}
    \caption{\textbf{Quantum Walk:} (a) case of unbiased coin; (b) case of biased coin with $r=0.6$ and $\ell=0.45$;  Cases of impurity M$_1$ in Eq.~\eqref{modA} for $\gamma_\text{imp}=6\gamma_\text{c}$ in (c), $\gamma_\text{imp}=\gamma_\text{c}$ in (d) and $\gamma_\text{imp}=\gamma_\text{c}/10$ in (e) with $\gamma_\text{c}=\ell-r$. Cases of impurity M$_2$ in Eq.~\eqref{modB}: $\phi=\pi/20$ in (f), $\phi=\pi$ in (g) and $\phi=\pi/2$ in (h). Cases of impurity M$_3$ in Eq.~\eqref{modC} for $\gamma_\text{L}=\gamma_\text{L}^\text{c}$, $\gamma_\text{R}=\gamma_\text{R}^\text{c}$ in (i), $\gamma_\text{L}=13\gamma_\text{L}^\text{c}/10$, $\gamma_\text{R}=\gamma_\text{R}^\text{c}/10$ in (l) and $\gamma_\text{L}=\gamma_\text{L}^\text{c}/10$, $\gamma_\text{R}=\gamma_\text{R}^\text{c}/1000$ in (m). Respective critical values read $\gamma_\text{L}^\text{c}=1.1 r$ and $\gamma_\text{R}^\text{c}=0.99 \ell$. For the panels (c) - (m), the impurity is placed at $n_\text{imp}=25$.
    }
    \label{fig4}
\end{figure*}

The Hatano–Nelson model can be effectively mapped onto a non-Hermitian DTQW. In this correspondence, the Hamiltonian’s non-reciprocal hopping amplitudes are encoded in the coin or shift operations of the quantum walk, thereby reproducing the essential features of biased and non-Hermitian dynamics. We adopt the following mapping, where the coin operator is defined as
%
%
\begin{equation}\label{coin}
    C=\begin{pmatrix}
        \sqrt{r} & \sqrt{1-r} \\
        \sqrt{1-\ell} & -\sqrt{\ell}
    \end{pmatrix},
\end{equation}
%
%
in the left/right coin basis, with $r$ and $\ell$ representing the transition probabilities for the walker to move to the right and left, respectively, satisfying $r,\ell \in [0,1]$. These coin parameters can be identified with the hopping amplitudes in the Hatano–Nelson model via the associations $\tR \sim \ell$ and $\tL \sim r$, such that the non-reciprocal character of the original model is captured through asymmetry in the quantum walk’s internal dynamics. In the continuum limit, this asymmetry manifests as a biased derivative term, representing a net drift induced by the imbalance, thus mirroring the non-Hermitian transport behavior characteristic of the Hatano–Nelson model~\cite{Childs_2009}.

The presence of an impurity in a DTQW can significantly alter the interference landscape and thus reshape the evolution of the walker. Several schemes exist for implementing such impurities, all premised on locally modifying the coin operator introduced in Eq.~\eqref{coin}. In the most straightforward scenario, we apply a scaling factor $\gamma_{\mathrm{imp}}$ exclusively at the designated impurity location:
%
%
\begin{equation}\label{modA}
    C_\text{imp}^\text{M$_1$}=\gamma_\text{imp}\, C \delta_{n,n_\text{imp}},
\end{equation}
%
%
with $n_\text{imp}$ the position of the impurity and  $\gamma_\text{imp}\in[0,1]$.~\cite{da_S_Teles_2021} 
This implementation directly modifies the walker time evolution, transforming it into a position-dependent one. The mismatch in coin parameters acts like a scattering center. A walker incident on the impurity can be partially reflected and transmitted, much like encountering a potential barrier.  Depending on the degree of mismatch, localized states may form near the impurity. These localized modes can lead to resonant scattering, where the walker’s amplitude becomes temporarily trapped.
Due to the non-unitary DTQW and its inherently non-reciprocal nature $r\neq\ell$, the impurity can accentuate directional biases in propagation. 
The case where $\gamma_{\text{imp}} \approx 1$ represents a negligible impurity, meaning the walker evolves nearly undisturbed. In contrast, when $\gamma_{\text{imp}} \approx 0$, the impurity acts as a strong barrier, effectively splitting the Hilbert space into two disconnected regions: one accessible to the left of the impurity and another beyond it. In the extreme limit $\gamma_{\text{imp}} = 0$, the walker is completely blocked from accessing the downstream region. Interestingly, when $\gamma_{\text{imp}} \approx \ell-r$, destructive interference at the impurity can give rise to a mode that remains dynamically localized at the defect site throughout the evolution---this behavior mirrors the impurity-ICSE observed in the Hatano–Nelson model.

A different approach to include the presence of the impurity consists in multiplying the coin operator in Eq.~\eqref{coin} by a phase factor $\text{e}^{\text{i}\phi}$:
%
%
\begin{equation}\label{modB}
    C_\text{imp}^\text{M$_2$}=\text{e}^{\text{i}\phi} C \delta_{n,n_\text{imp}},
\end{equation}
%
%
with $\phi\in[0,2\pi)$.~\cite{Wojcik_2012,Li_2013,Zhang_2014}
The extra phase modifies the interference pattern of the DTQW. This can result in shifts in the positions of constructive and destructive interference, altering the probability distribution of the walker, and can additionally modify interference pathways in the DTQWca. While its magnitude remains unity --- unlike the amplitude-defect case in Eq.~\eqref{modA} --- a phase-only defect cannot create a strong barrier that fragments the walker's Hilbert space. Instead, it affects walker dynamics by altering interference patterns. Specifically, when the phase parameter is $\phi \approx 0$, the impurity is effectively weak; at $\phi \approx \pi/2$, it induces strong \emph{phase} scattering yet does not fragment the network; and at $\phi = \pi$, it produces behavior reminiscent of the ICSE, with a localized mode forming due to destructive interference at the defect. However, none of these regimes result in Hilbert-space fragmentation.

Finally, we can modify the coin operator to include an asymmetric gain or loss at the impurity site.~\cite{Mochizuki_2016} This is obtained by adjusting the coin operator at the impurity site as
%
%
\begin{equation}\label{modC}
    C_\text{imp}^\text{M$_3$}=\begin{pmatrix}
        \sqrt{\gamma_\text{R}} & 0 \\
        0 & \sqrt{\gamma_\text{L}}
    \end{pmatrix}\cdot C \delta_{n,n_\text{imp}},
\end{equation}
%
%
where $\gamma_\text{R}$ and  $\gamma_\text{L}$ are gain/loss factors with one possibly greater than one and the other less than 1. In turn, motion to the right might be amplified (if $\gamma_\text{R}>1$) while the motion to the left is attenuated (if $\gamma_\text{L}<1$), i.e., the procedure accentuates the inherent non-reciprocity of the Hatano-Nelson model. Also, this model of impurity as the one in Eq.~\eqref{modA} allows for Hilbert-space fragmentation when the gain/loss factors $\gamma_\text{R}$ and $\gamma_\text{L}$ are both small and highly imbalanced. If their ratio remains close to unity, the impurity remains weak. However, an extreme imbalance causes the impurity to act as a strong barrier, effectively dividing the walker's accessible regions --- at the limit, the walker cannot cross into the downstream fragment. Interestingly, when the ratio satisfies
$\frac{\gamma_\text{R}}{\gamma_\text{L}} = \frac{\ell}{r}$,
destructive interference at the defect site generates a dynamically localized mode— a hallmark of the ICSE.

In Fig.~\ref{fig4}, we present a comprehensive visualization of the DTQW phenomena discussed throughout this work. In Fig.~\ref{fig4}(a), we present the case of the unbiased coin, where the walker develops the characteristic probability bimodal distribution during time evolution. In Fig.~\ref{fig4}(b), we present the case of the biased coin with $r/\ell\approx 0.75$, in which the bimodal symmetry is disrupted due to directional bias. In Figs.~\ref{fig4}(c)-\ref{fig4}(e), we present the case with a defect of type M$_1$ for three different strengths of the impurity $\gamma_\text{imp}$: weak, intermediate, and strong. Analogously, in Figs.~\ref{fig4}(f)-\ref{fig4}(h), we present results for the case of a defect of type M$_2$ for three different values of the phase factor characterizing this model. Finally, in Figs.~\ref{fig4}(i)-\ref{fig4}(m), we present results for the last model of disorder M$_3$, considering three different cases for the gain and loss coefficients. 
Crucially, in each set the middle panel exemplifies the ICSE-\emph{like} regime, where a walker mode becomes persistently localized near the impurity site, coexisting with bulk drift. This robust localized behavior is the DTQW counterpart to the ICSE described earlier.

\section{Conclusions}\label{sec_iv}
In this manuscript, we have studied the spectral properties of the Hatano-Nelson model with a generic impurity's position and strength. We have shown analytical solutions for both the periodic and open boundary conditions. For both cases, the associated wavevectors obey a transcendental constraint.

Interestingly, we have found that the strong impurity limit fixes the matching condition to open boundary conditions, causing the chain to fragment. Subsequently, the non-Hermitian skin effect emerges naturally under periodic boundary conditions, confirming earlier studies.~\cite{Roccati_2021, Liu2020} Under open boundary conditions, we uncovered that the impurity introduces a second length scale capable of competing against the non-Hermitian skin effect. This is explicit in the case of an infinitely strong impurity, where one mode (the impurity mode) is trapped at the impurity site despite the model's non-reciprocal hopping. For moderate impurity strength, this mode decays exponentially away from the impurity.

We investigated the parameter regions where the impurity dominates the impurity mode over the non-Hermitian skin effect. At the interface between the two regimes, we found the impurity-induced counter-skin effect where the impurity mode becomes constant within one pristine subchain. We provided numerical and analytical evidence and demonstrated that the impurity-induced counter-skin effect may be identified from the non-Hermitian skin effect. We have verified the appearance of the ICSE in multi-band non-Hermitian models --- see SM.

Our study may serve as a blueprint for the multi-impurity case. Supposing that they are placed at distinct sites $l_1,\dots, l_n$ and have strengths $\delta_1,\ldots,\delta_n$, each impurity may induce its own localization strength in case a domain wall architecture ($\delta_1=\ldots=\delta_n$ for neighboring sites) is prevented. The argument is obvious from the case $n=2$, since $\vert \delta_1\vert \rightarrow \infty$ fragments the chain, one of which contains the second impurity. Now, $\vert \delta_2\vert \rightarrow \infty$ splits the fragment into pieces, traps a second state at $l_2$ and the argument may be continued for further impurities. 

There are two conclusions to be drawn from this. Firstly, placing an impurity on every second site is sufficient to isolate all sites from their respective neighbors for sufficient impurity strength. In turn, the NHSE vanishes.

Secondly, and following our discussion of the out-of-band transition for a single impurity, placing one at every site with sufficient strength, and at best with alternating signs, effectively destroys the band structure of the chain. Although the dispersion relation $E(kd)$ from the pristine case can be used together with a proper quantization condition, the state's energy is determined mainly by the local impurity, i.e., energy eigenvalues appear fully misplaced with respect to the dispersion relation due to generally complex wavevectors.

In the last part of this work, we have explored how the ICSE can manifest in a DTQW by incorporating three distinct impurity models --- amplitude (M1), phase (M2), and gain/loss (M3) modifications. A central contrast between the DTQW and the original Hatano–Nelson framework lies in the nature of the ICSE manifestation: in the Hatano–Nelson model, the ICSE appears as a localized impurity eigenmode that can be isolated in the system’s spectrum, whereas in the DTQW, the inherently non-Hermitian, dynamic evolution precludes such static spectral isolation. Instead, one must scrutinize the full-time evolution of the walker's state ensemble to detect ICSE-like behavior. We argue that a reliable signature in this context is the emergence of a component that remains persistently localized around the impurity site, despite an overall biased drift. The sustained localization of such a state thus serves as a practical indicator of interference-driven cancellation of the biased coin and the realization of ICSE dynamics within the DTQW framework.

\bibliography{bibliography}

\section*{Acknowledgement}
We acknowledge interesting and fruitful discussions with Geza Giedke, Flore Kunst, Thomas Frederiksen, Shayan Edalatmanesh, and Carolina Martinez Strasser. This research was funded by the IKUR Strategy under the collaboration agreement between the Ikerbasque Foundation and DIPC on behalf of the Department of Education of the Basque Government. D.B. acknowledges the support from the Spanish MICINN-AEI through Project No.~PID2020-120614GB-I00~(ENACT)d, the Transnational Common Laboratory $Quantum-ChemPhys$ and the Department of Education of the Basque Government through the project PIB~2023~1~0007 (STRAINER).
\section*{AUTHOR DECLARATIONS}
\subsection*{Conflict of Interest}
The authors have no conflicts to disclose.
\subsection*{Author Contributions}

\comment{Conceptualization (NL,DB); Data curation (NL,DB); Formal analysis (NL,DB); Funding acquisition (NL,DB); Investigation (NL,DB); Software (NL,DB); Visualization (NL,DB); Writing – original draft (NL,DB); Writing – review \& editing (NL,DB).}

Both NL and DB contributed equally to all aspects of this research, including conceptualization, data curation, formal analysis, funding acquisition, investigation, software development, visualization, and the writing of both the original draft and subsequent revisions.

\section*{DATA AVAILABILITY}
The data associated with the figures are in this work are available via Zenodo at the following URL. The codes can be shared upon reasonable request.

\onecolumngrid

\pagebreak

\renewcommand{\theequation}{SE\arabic{equation}}
\renewcommand{\thefigure}{SF\arabic{figure}}
\renewcommand{\bibnumfmt}[1]{[#1]}
\renewcommand{\citenumfont}[1]{#1}
\setcounter{equation}{0}
\setcounter{figure}{0}
\setcounter{table}{0}
\setcounter{section}{0}

\begin{center}
   \large{\textbf{SUPPLEMENTAL MATERIAL FOR:}} \\ 
    \vspace{.5cm}
   \Large{Impurity-induced counter skin-effect and linear modes\\ in the Hatano-Nelson model} \\
   \vspace{0.5cm}
   \normalsize{Nico Leumer and Dario Bercioux}
\end{center}

\section{Eigenvector equation}
\subsection{Solution under Open Boundary Conditions}
\label{section_solOBC}
The Hamiltonian density $\mathcal{H}_l$ from Eq.~\eqref{eq: Hamiltonian} in real space $\vec{\Psi} = (c_1,\ldots,\,c_N)^\mathrm{T}$, satisfying $\hat{H}_l = \vec{\Psi} \mathcal{H}_l\vec{\Psi}^\dagger$, motivates the choice $\psiR=(\alpha_1,\,\ldots,\,\alpha_{l-1}, \beta, \,\gamma_{l+1},\,\ldots,\,\gamma_N)^\mathrm{T}$ with $\alpha_j$ ($\gamma_j$) as entries before (after) the impurity and $\beta$ at the impurity site $l$ in analogy to a scattering ansatz in textbook quantum mechanics.~\cite{griffiths2018} Then, the rhs eigenvector equation $\mathcal{H}_l\,\psiR = E\,\psiR$ under OBC yields straightforwardly the recursive formulae
\begin{subequations}\label{eq_12}
\begin{align}
    \label{eq: Fib alpha}
    \tL \alpha_{j+1} &= E\alpha_j -\tR\,\alpha_{j-1},\quad j=2,\ldots\,l-2\\
    \label{eq: Fib gamma}
    \tL \gamma_{j+1} &= E\gamma_j -\tR\,\gamma_{j-1},\quad j=l+2,\ldots\,N-1
\end{align}
\end{subequations}
for the bulk, the OBC in its naive form
\begin{subequations}\label{eq_13}
    \begin{align}
    \label{eq: OBCL, naive form}
    \tL \alpha_{2} &= E \alpha_1,\\
    \label{eq: OBCR, naive form}
    0 &= E \gamma_N-\tR \gamma_{N-1},
\end{align}
\end{subequations}
matching conditions at the impurity as
\begin{subequations}\label{eq_14}
    \begin{align}
    \label{eq: MC1}
    \tL \beta &=  E\alpha_{l-1} -\tR\,\alpha_{l-2},\\
    \label{eq: MC2}
    \tL \gamma_{l+1} &=  (E-\delta)\beta -\tR\,\alpha_{l-1},\\
    \label{eq: MC3}
    \tL \gamma_{l+2} &= E \gamma_{l+1} -\tR\beta.
\end{align}
\end{subequations}
Notice that the respective constraints for lhs eigenvectors follow by mutual exchange of $\tL$ and $\tR$. Although $\ldots,\,\alpha_{-1},\,\alpha_0,\,\alpha_l,\,\alpha_{l+1} ,\ldots$ (and similar for $\gamma$) are never part of the eigenvector $\psiR$, we define those terms as the natural continuation of Eqs.~\eqref{eq_12} beyond their initial limits. In turn, this simplifies the OBC $\alpha_0 = \gamma_{N+1} =0$ as long as $\tL \tR \neq 0$. Similarly, matching constraints Eqs.~\eqref{eq: MC1}, \eqref{eq: MC3} state the continuity condition
\begin{align}\label{eq: continuity conditon}
   \beta = \alpha_l=\gamma_l. 
\end{align}

Solutions to $\alpha_j$, $\gamma_j$ follow from the standard ansatz~\cite{Hoggatt-1974, Webb-1969} $\alpha_j,\gamma_j\propto r^j$ ($r\neq 0$) transforming Eqs.~\eqref{eq_12} into a quadratic equation in $r$ and its two roots read $2\,r_\pm  = x\pm \sqrt{x^2-4y}$, where $x= E/\tL$, $y = \tR/\tL$ for shortness. In turn, $\alpha_j$, $\gamma_j$ follow as superpositions of $r_\pm^j$ and coefficients can be determined from initial values, e.g., $\alpha_{0,1}$ ($\gamma_{N,N+1}$). Realizing that also $\Fk{j}=(r_+^j-r_-^j)/(r_+-r_-)$ (with $\Fk{0} = 0$, $\Fk{1} = 1$, $\Fk{-1} = -1/y$) is a solution too, one may construct\cite{Leumer21}

\begin{align}
    \alpha_j  &= a_1 \Fk{j} + a_0 \Fk{j-1},\\
     \gamma_j &= c_1 \Fk{j-N} + c_0 \Fk{j-N-1}.
\end{align}
with some coefficients $a_{0,1}$, $c_{0,1}$. For $j=0,1$, we identify $a_1=\alpha_1$, $a_0 =-\alpha_0 y$ (and similar for $\gamma_j$) such that the OBC sets $a_0 = c_0 = 0$. The matching constraint $\alpha_l = \gamma_l$ relates $a_1$, $c_1$ such that we arrive at
\begin{subequations}\label{eq5}
\begin{align}
   \label{eq: alpha's, delta finite OBC}
   \frac{\alpha_j}{\alpha_1} &= \Fk{j},\\
   \label{eq: gamma's, delta finite OBC}
     \frac{\gamma_j}{\alpha_1} &=\frac{\mathcal{F}_k(l)}{\mathcal{F}_k(l-N-1)}~\mathcal{F}_k(j-N-1)
\end{align}
\end{subequations}
with $\alpha_1$ as the normalization constant of $\psiR$. Since the matching constraint fixed $\beta=\alpha_l = \gamma_l$, only the energy $E$ remains to be fixed. 

More convenient is to introduce wavevectors $k \in \mathbbm{C}$ and a dispersion relation $E\equiv E(k)$. The latter may be chosen as $x \eqcolon 2\sqrt{y}\cos(kd)$, with $d$ as the lattice constant, which gives Eq.~\eqref{spectrum_OBC} after replacing $x= E/\tL$, $y = \tR/\tL$. This peculiar choice allows us to transform $r_\pm = \sqrt{\tR/\tL} e^{\pm \ii kd}$ and in turn 
\begin{align}\label{eq: Fk in kd form}
    \Fk{j} = \sqrt{\frac{\tR}{\tL}}^{j-1} \frac{\sin(kd\,j)}{\sin(kd)}
\end{align}
simplifies. Imposing Eqs.~\eqref{eq5}, and \eqref{eq: Fk in kd form} into the last remaining constraint, that is Eq.~\eqref{eq: MC2}, one finds the quantization condition displayed in Eq.~\eqref{eq_two}. Notice that adapted schemes also exist for systems including (effective) next nearest neighbor hoppings~\cite{Leumer21, Leumer2023, Leumer-2020, Edvardsson2022} where the recursion range is enlarged.
\subsubsection{Solver dependent issues and numerical verification}
%
%
%
%
%
%
\begin{figure}
    \centering
    \includegraphics[width = 0.98\columnwidth]{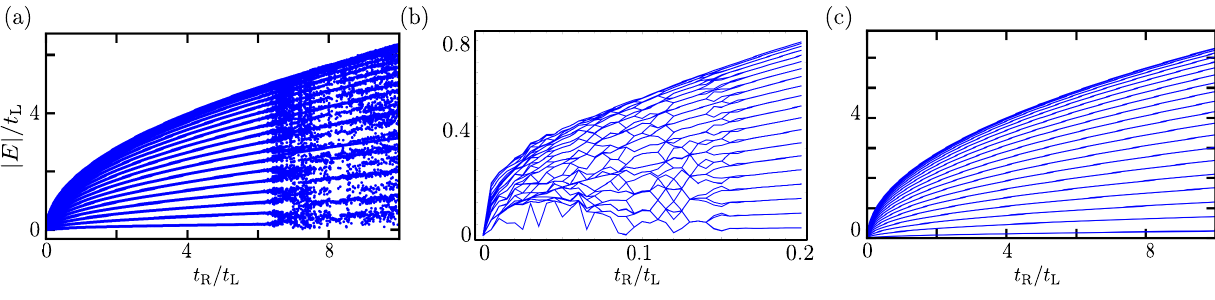}
    \vspace{-0.4cm}
    \caption{\textbf{Comparison of numerical and analytical methods.} Solver-dependent numerical errors calculating the energy spectrum of the Hamiltonian from Eq.~\eqref{eq: Hamiltonian} in the main text under OBC and at $\delta = 0$, $N=40$. (a) Data obtained from numpy's linear algebra routines.~\cite{harris_2020} (b) Numerical diagonalization with Wolfram Mathematica\textregistered\,  using machine precision.~\cite{Mathematica} (c) Analytical energy spectrum obtained numerically with the Fibonacci polynomials.}
    \label{FigSM1}
\end{figure}
%
%
%
%
\begin{figure}
    \centering
    \includegraphics[width = \columnwidth]{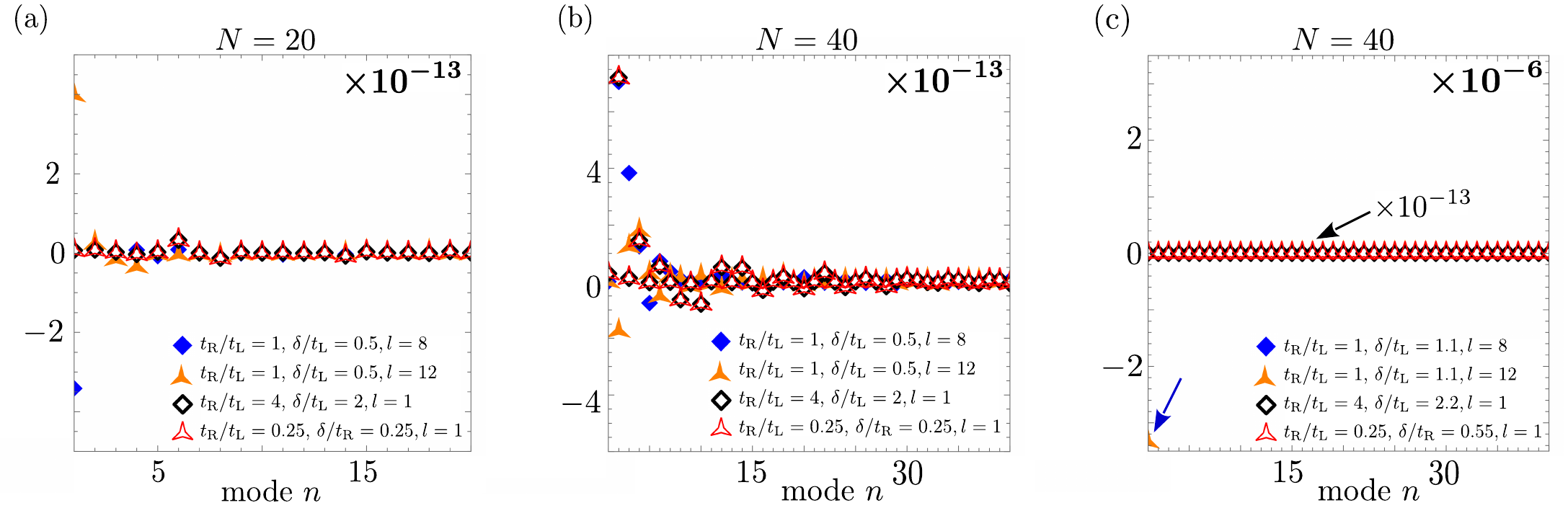}
    \vspace{-0.4cm}
    \caption{\textbf{Numerical confirmation of the quantization condition.} (a) ((b)) Difference between left/right side of Eq.~\eqref{eq_two} is of the order $10^{-13}$ for $N=20$ ($N=40$). (c) Larger impurity strength $\delta$ may cause a drop in accuracy, here for a single value to $10^{-6}$.}
    \label{FigSM2}
\end{figure}
%
%
For the purpose of illustration, in Figs.~\ref{FigSM1}\R{a} and~\ref{FigSM1}\R{b}, we show data obtained with machine precision using standard Python libraries~\cite{harris_2020} (Wolfram Mathematica\textregistered~\cite{Mathematica}) as solver next to the proper results in Fig~\ref{FigSM1}\R{c}. Apparently, errors in the spectrum are solver-dependent.

In contrast, we show the \emph{numerical} satisfaction of Eq.~\eqref{eq_two} (with an overhaul scale of $10^{-13}$) for the vertical axis in n Figs.~\ref{FigSM2}\R{a}, and~\ref{FigSM2}b. Data points illustrate the difference between the left and right side of the quantization condition for each energy obtained from Eq.~\eqref{eq: Hamiltonian}. The spectrum is calculated in exact terms, yet associated wavevectors inherit machine precision due to Eq.~\eqref{spectrum_OBC}. In Fig.~\ref{FigSM2}\R{c}, we show that this may cause significant drops in accuracy simply by changing the model's parameter. 

Overhaul, even a pure numerical treatment with potentially large mistakes may benefit from the quantization condition since Eq.~\eqref{eq_two} allows a precise quantification of the errors made. Then Eqs.~\eqref{eq5} provide results for the eigenvectors or the NHSE, with known quantified accuracy even when exact analytic routines are out of scope due to large system sizes~etc.

\subsubsection{Limit case $|\delta|\to\infty$}
The earlier analogy to scattering theory can be expanded to the tunneling effect. Within energetically forbidden regions $E<V(x)$, wavefunctions experience an exponential decay such that the barrier becomes fully non-transparent when $V(x)\rightarrow \infty$.~\cite{griffiths2018} In full analogy, the chain separates into three fully independent fragments (despite the fact that $\tLR\neq 0$ remains finite) as can be analytically anticipated from the eigenvector equation. Recalling that $\alpha_j$, $\gamma_j$, $\beta$, $k\equiv k_\delta$ depend on the impurity strength, one has merely to properly cope with the infinity to ensure normalizable solutions. The key role is adopted by Eq.~\eqref{eq: MC2} from where we extract two distinct behaviors; either (i) we have that $E\neq \delta$ but $\beta\rightarrow 0$ vanishes faster than $\vert \delta\vert\rightarrow \infty$ grows or (ii) $E = \delta$ and $\beta = 1$.

\paragraph{Scenario (i)} The condition $\beta\rightarrow 0$ yields $\alpha_l\rightarrow0$ and $\gamma_l\rightarrow0$ (cf.~Eq.~\eqref{eq: continuity conditon}), i.e. imitating OBC at the impurity's position. In turn, $N-1$ modes behave as if the chain was interrupted at $j=l$ as sketched in Fig.~\ref{Fig1}\R{e}. Generally, the fragments have a non-degenerate energy profile due to $\sin(kd\,l)\rightarrow 0$, $\sin[kd(N-l+1)]\rightarrow 0$, i.e., the constraints $\alpha_0 =0$, $\alpha_l\rightarrow0$ and $\gamma_l\rightarrow0$, $\gamma_{N+1}=0$ are not compatible for the same eigenvector. In turn, one has that either all $\alpha_j =0$ or all $\gamma_j=0$, signaling the ``splitting of the chain''.

\paragraph{Scenario (ii)} This case concerns the impurity mode \imp alone. Inserting $E = \delta$, $\beta = 1$ into Eqs.~\eqref{eq_12}-\eqref{eq: MC3} demands that $\alpha_1,\,\ldots,\,\alpha_{l-1}$, $\gamma_{l+1},\,\ldots,\,\gamma_{N}\rightarrow 0$ diminish all faster than impurity strength increases. Hence, this mode is trapped at the impurity and has an associated energy of $\vert E\vert \rightarrow \infty$. In case of finite (but dominant $\delta$) \imp is exponentially localized around the impurity, i.e. its energy is only about $\delta$ due to the hybridization with neighboring sites. 
\subsection{Solution under Periodic Boundary Conditions}
For PBC, we can still apply the same strategy on $\mathcal{H}_l\,\psiR = E\,\psiR$ as in section \ref{section_solOBC} before. Only the boundary condition naturally changed to
\begin{subequations}\label{eq_18}
    \begin{align}
    \tL \alpha_2 &= E \alpha_1 - \tR \gamma_N,\\
    \tL \alpha_1 &= E \gamma_N -\tR \gamma_{N-1}.
\end{align}
\end{subequations}
The extension of the recursive sequence beyond its natural limits of the eigenvector equation yields now $\alpha_0 = \gamma_N$, $\alpha_1 = \gamma_{N+1}$ provided that $\tL\tR\neq 0$. In terms of $r_\pm$, we construct $\alpha_j = a\,r_+^j + b\,r_-^j$, $\gamma_j = c r_+^{j-N} +d r_-^{j-N}$ and PBCs imply $a = c$, $b=d$. After some algebra, Eqs.~\eqref{eq: MC2}, \eqref{eq: continuity conditon} yield 

\begin{align}\label{eq: quantization constraint with delta PBC}
    \frac{\delta}{\tL} \frac{r_+^N-r_-^N}{r_+-r_-} = -\left(1-r_+^N\right)\left(1-r_-^N\right),
\end{align}
the quantization condition only before the substitution of $r_\pm$. 

\paragraph{Case of $\delta = 0$:} Since $r_+$ is generally a complex quantity, we may introduce a wavevector $q$ via the polar form 
\begin{align}\label{eq: r_+ polar form}
    r_+ = \vert r_+\vert e^{\ii qd},\quad qd\in \mathbbm{R}.
\end{align}
At $\delta = 0$, the roles of $r_\pm $ in Eq.~\eqref{eq: quantization constraint with delta PBC} are mutual exchangeable since $r_+ r_- = \tR/\tL$ is a general property. Focusing on $r_+$, we find $ \vert r_+\vert = 1$ and $qd = 2\pi n/N$ with $n=1,\,\ldots,\,N$. As expected, the dispersion relation $E=E(q)$ is the one from PBC, i.e., Eq.~\eqref{spectrum_PBC}, and it follows from Eq.~\eqref{eq: r_+ polar form} and $2r_+ = x+\sqrt{x^2-4y}$ with $x=E/\tL$, $y=\tR/\tL$.

Of course, any other choice for a dispersion relation is also possible. However, an unwise choice comes at the price that Eq.~\eqref{eq: quantization constraint with delta PBC} adopts an unnecessarily complicated structure. For instance, with Eq.~\eqref{spectrum_OBC}, i.e., $r_\pm = \sqrt{\tR/\tL} e^{\pm \ii kd}$, Eq.~\eqref{eq: quantization constraint with delta PBC} turns into
\begin{align}\label{eq: PBC+ zero delta, quantization for kd}
    \cos(kd N) = \frac{t_{\rm L}^N+t_{\rm R}^N}{2\sqrt{\tL \tR}^N}.
\end{align}
The reason for this is that now $kd$ has to cancel the NHSE effect from $r_\pm$'s prefactor $\sqrt{\tR/\tL}$.

\paragraph{Case of $\vert\delta\vert \rightarrow \infty$:} Now, the quantization constraint from Eq.~\eqref{eq: quantization constraint with delta PBC} becomes $\sin(kdN) = 0$ with substitution $r_\pm = \sqrt{\tR/\tL} e^{\pm \ii kd}$. This manifests that the model under PBC fragments into a pristine Hatano-Nelson chain under OBC, with energy eigenvalues follow from Eq.~(\R{3}) in the main text and wavevectors $k_nd = n\pi/N$, $n=1,\,\ldots,\, N-1$, and the isolated impurity site. Similarly to case discussed in the main text, the OBC for $\vert\delta\vert \rightarrow \infty$ manifests at the impurity position. Notice here, that Eq.~\eqref{eq: quantization constraint with delta PBC} was in fact independent of the impurity position $l$, since former PBC allows to relabel all sites arbitrarily.
\section{ICSE}
\subsection{Energy dependency}
\begin{figure*}[ht]
    \centering
    \includegraphics[width = \textwidth]{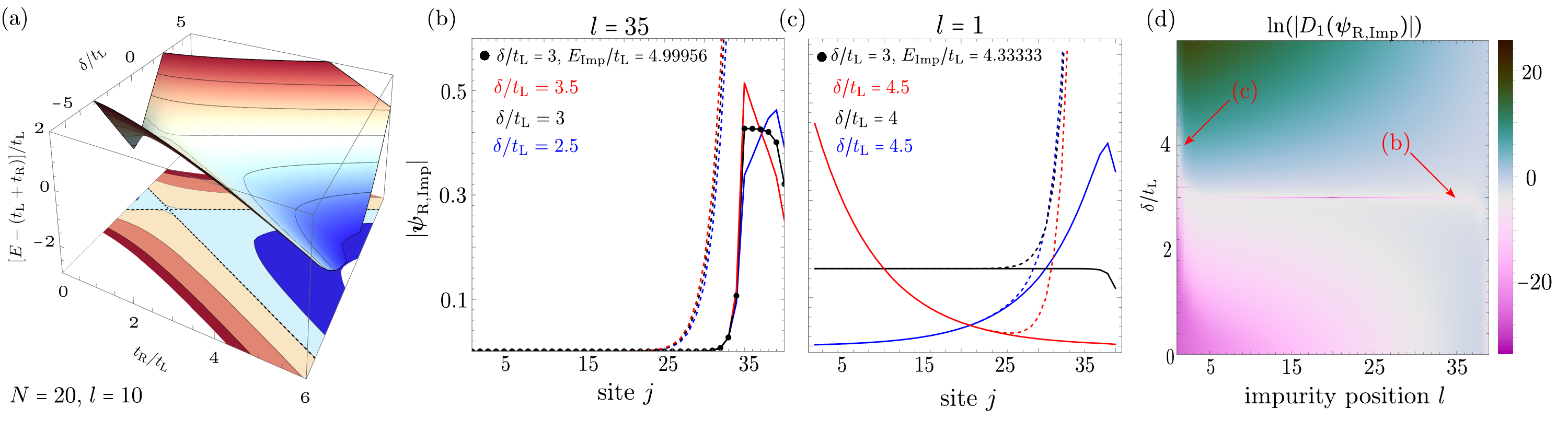}\vspace{-0.3cm}
    \caption{\textbf{Impurity-induced counter skin-effect, energy and parameter dependence.} (a) Top: Maxima of the absolute energy values $E_{\rm Imp}$ shifted by $-(\tL-\tR)$. Bottom: Contour plot. Red (blue) colors indicate positive (negative) $E-(\tL-\tR)$. Black dashed lines highlight $E=\tL-\tR$ as a guide to the eye. (b) $\boldsymbol{\psi}_{\rm R, Imp}$ for various impurity strength's $\delta$. The analytic approximation (black dots) for the ICSE (solid black) at $\DICSE = \tR-\tL$ confirms the data from exact analytic diagonalization. (c) For terminal positions $l$, the ICSE requires a modification of $\DICSE$. (d) Dependence of ICSE (central purple line) on $\delta$ and $l$. The lower left corner (purple) corresponds to exponentially localized states. In panels (b)-(d), we set $N=40$ and $\tR/\tL = 4$. Discussion in the main text. All computational data is obtained by exact diagonalization.}
    \label{FigSM3}
\end{figure*}

The energy dependency of the impurity mode under the ICSE can be approximated analytically directly from the eigenvector equation 
for impurities placed in the bulk. As ansatz, we recall that the energy of \imp is about $\delta$, i.e., we set $E_{\rm imp} = \delta + \epsilon(\tL, \tR, \delta)$. Here, 
$\epsilon$ accounts for the hybridization of \imp with neighboring sites and its value is typically unknown. In the following, we approximate $\epsilon$, exploiting that \imp adopts a nearly constant profile (to either side of the impurity) in the case of the ICSE. The spatial behavior of \imp is encoded in Eqs.~\eqref{eq: Fib alpha},~\eqref{eq: Fib gamma} and since both are equivalent under renaming $\alpha\rightarrow\gamma$, we focus on Eq.~\eqref{eq: Fib gamma}. Also, the spectrum changes only a sign for $\delta\rightarrow -\delta$, hence we assume  $\delta>0$, i.e. that $\delta = \tR-\tL>0$ in Eq.~\eqref{eq: delta ICSE}. Inserting $E$, $\delta$ into Eq.~\eqref{eq: Fib gamma} and demanding that $\gamma_{j} = \gamma_{j-1}$, grants
\begin{align}
    \gamma_{j+1} = \frac{\epsilon-\tL}{\tL} \gamma_{j}.
\end{align}
Since the prefactor needs to be one for a constant profile, we find $\epsilon = 2\tL$ and therefore $E = \tL+\tR$. Although not exact, the approximated energy is rather close to the true value such that the ICSE can be reproduced from Eq.~\eqref{eq5} with $kd = \arccos[(\tL+\tR)/\sqrt{4\tL\tR}]$ from Eq.~\eqref{spectrum_OBC}. For comparison of \imp with numerical data, we refer to the black dots in Fig.~\ref{Fig1}\R{h} and Fig.~\ref{FigSM3}\R{b}.

In Fig.~\ref{FigSM3}\R{a}, we show the numerical energy spectrum obtained with infinite precision. Indeed, with $\delta$ fixed by Eq.~\eqref{eq: delta ICSE}, the energy is $E = \tL+\tR$ (dashed 
lines). The only exception is when the NHSE becomes suppressed for $\tL\approx\tR$ as expected. 

In Fig.~\ref{FigSM3}\R{b}, we show \imp for various values of $\delta$ and $N=40$, $l=35$. For the ICSE (black), \imp shows a flat profile behind the impurity and its energy $E_{\rm imp}/\tL =  4.99$ is close to the approximated value $E\tL = 5$ for $\tR/\tL = 4$.
%
%
%
%
%
%
%
%
\subsection{Intersection points and $y_{\rm c,\pm}$}
The values for $\tR/\tL = y_{\rm c,\pm}$ follow from the intersection points (cf. blue/orange in dots Fig.~\ref{Fig3}\R{c}) when $\delta = \pm (\tR-\tL)$ equals $\dcrit$ in Eq.~\eqref{eq_dcrit} from the out-of-band transition. We find
\begin{align}\label{eq: ycrit}
    y_{\rm c, \pm} = \frac{\left(b\pm \sqrt{b^2+4}\right)^2}{4},\qquad b\coloneqq \frac{N+1}{l\,(N+1-l)}
\end{align}
and $b\le 1 +\nicefrac{1}{N}$ (cf.~Fig.~\ref{Fig2}\R{c}) implies that $y_{\rm c, +}\ge 1\ge y_{\rm c, -}$.
\subsection{Impurities at edge positions}
In Fig.~\ref{FigSM3}\R{c}, we placed the impurity at the edge, i.e., $l=1$. Varying the value of $\delta$ and observing the spatial profile of \imp shows clearly that the ICSE (black) is present. While for bulk placements, we anticipate $\delta/\tL=3$ (cf. Eq.~\eqref{eq: delta ICSE}), the required value $\delta/\tL=4$ is actually larger. Similarly, the energy $E_{\rm imp}$ also differs. Using the gradient from Eq.~\eqref{eq: definition Dx}, Fig.~\ref{FigSM3}\R{d} confirms that impurities closer to the edges demand an adaptation to  Eq.~\eqref{eq: delta ICSE}. Here, the small purple line in the center corresponds to the ICSE. 

\begin{figure}[ht]
    \centering
    \includegraphics[width = \columnwidth]{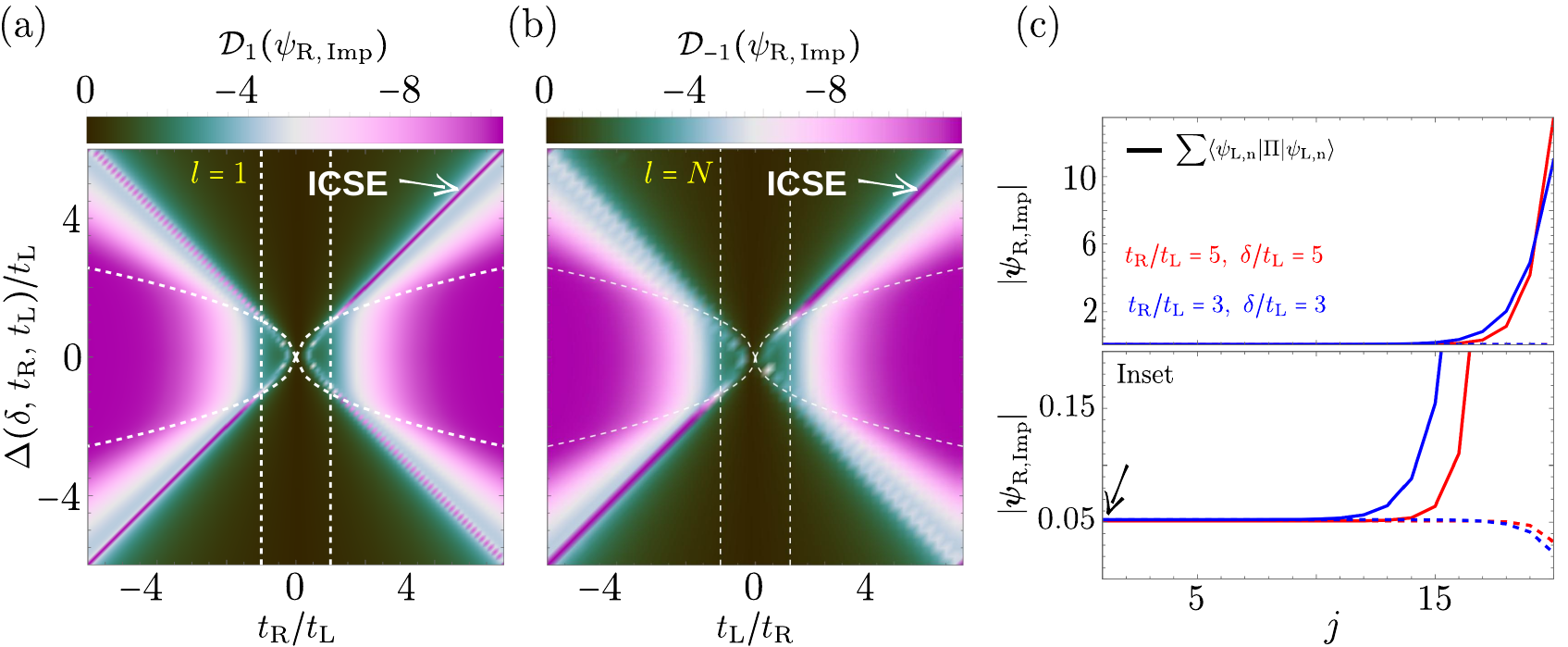}
    \caption{\textbf{ICSE phase diagram and ICSE detection in NSHE.} (a) The ICSE (purple lines) also exists for impurities on the first site $l=1$, but its strength demands an adaptation. (b) For impurities on the last site $l=N$, the ISCE exists as well. Notice the modified horizontal axis to better display the linear parameter dependence. (c) As a local quantity, the ICSE (dashed) may be visible from the NHSE (solid) upon closer inspection. All data from exact diagonalization for $N=20$, $l=1$ in (a) and $l=20$ in (b).}
    \label{FigSM4}
\end{figure}

We conducted a closer investigation in Fig.~\ref{FigSM4}\R{a} for $l=1$ and in Fig.~\ref{FigSM4}\R{b} for $l=N$. The shifted gradient $\mathcal{D}_{\pm 1}=\ln(10^{-5} + |D_{\pm1}(\psi_{\rm R,\, Imp})|)$ provides improved resolution and the ISCE resides on the separated white/purple lines. The vertical dashed lines mark $\vert \tR/\tL \vert = 1 $. When the impurity is placed on the first site (Fig.~\ref{FigSM4}\R{a}), the parameter constraint $\delta/\tL = \pm (\tR/\tL-1)$ from Eq.~\eqref{eq: delta ICSE} is falsified. Instead, the ICSE requires $\delta/\tL = \pm \tR/\tL$, i.e., $\delta = \pm \tR$, for sufficient $\tR/\tL$. In Fig.~\ref{FigSM4}\R{b}, we inverted the horizontal quantity $\tR/\tL$ into $\tL/\tR$ to display the linear trend $\delta/\tR = \pm \tL/\tR$, i.e., $\delta = \pm \tL$ for sufficient $\tL/\tR$.

Finally, in Fig.~\ref{FigSM4}\R{c}, we show that the ICSE may be observed from the NHSE.
\subsection{ICSE for multiband models: non-reciprocal SSH model}
\begin{figure}
    \centering
    \includegraphics[width=\linewidth]{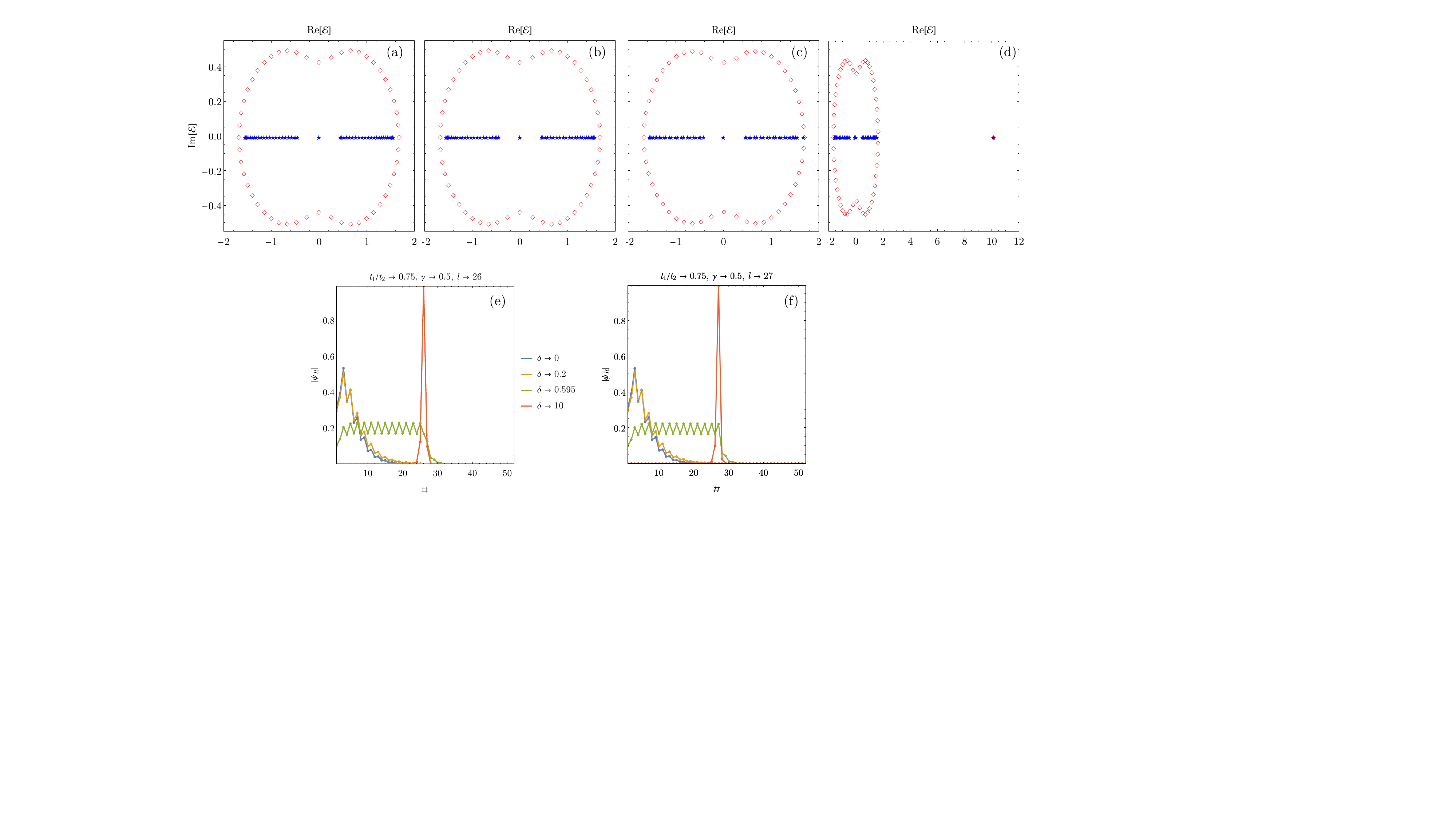}
    \caption{\textbf{ICSE in multiband system, topological regime.} Complex spectra for the non-reciprocal SSH model in the topological regime for various impurity strengths: (a) $\delta=0$, (b) $\delta =0.2t_2$, (c) $\delta=0.595 t_2$, and (d) $\delta=10t_2$. The presence of a state at zero energy characterizes the topological phase regime. Modulus of the wave function associated with the impurity $|\psi_\text{imp}|$ for the impurity in the 26th unit cell placed on the "A" site in (e) and "B" site in (f).  We have set $t_2=1$.}
    \label{fig_ICSE_SSH_Topo}
\end{figure}

\begin{figure}
    \centering
    \includegraphics[width=\linewidth]{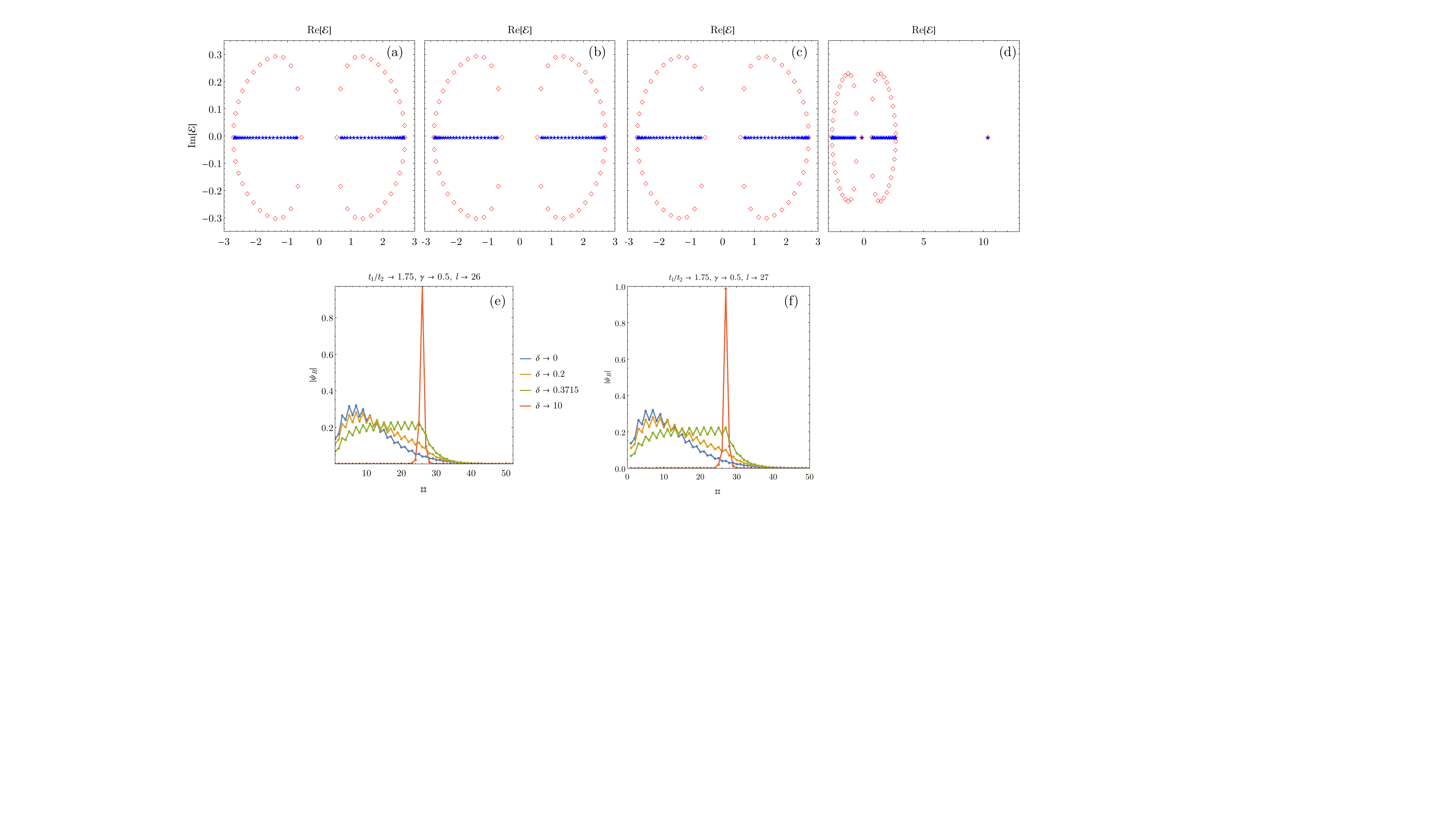}
    \caption{\textbf{ICSE in multiband system, trivial regime.} Complex spectra for the non-reciprocal SSH model in the trivial regime for various impurity strengths: (a) $\delta=0$, (b) $\delta =0.2t_2$, (c) $\delta=0.3715 t_2$, and (d) $\delta=10t_2$. The presence of an energy gap characterizes the trial phase regime. Modulus of the wave function associated with the impurity $|\psi_\text{imp}|$ for the impurity in the 26th unit cell placed on the "A" site in (e) and "B" site in (f). We have set $t_2=1$.}
    \label{fig_ICSE_SSH_Trivial}
\end{figure}

In this section, we prove the existence of the ICSE in a multi-band non-Hermitian system presenting NHSE. For this purpose, we consider the non-reciprocal SSH model.~\cite{Lieu_2018} This model is a 1D model with a base composed of the lattice sites "A" and "B", where we distinguish the intracell from the intercell hopping. The non-reciprocal character is introduced in the intracell hopping via a real parameter $\gamma$. The Hamiltonian of this model reads:
%
%
\begin{equation}
    \mathcal{H}_\text{NR-SSH}=\sum_n (t_1+\gamma)a_n^\dag b_n+ (t_{1}-\gamma)b_n^\dag a_n + t_{2}(a_{n+1}^\dag b_n+b_n^\dag a_{n+1}),
\end{equation}
%
%
where $t_1$ is the intra- and $t_2$ inter-cell hopping terms. Here $a_n^{(\dag)}$ is the annihilation (creation) operator for a state on the lattice site "A", whereas $b_n^{(\dag)}$ is the annihilation (creation) operator for a state on the lattice site "B". As for the case of the Hatano-Nelson model in Eq.~\ref{eq: Hamiltonian}, we include the possibility of having a defect of strength $\delta$ that can seat in the $n$th unit cell either on "A" or "B" sublattice. The equivalent Bloch Hamiltonian reads:
%
%
\begin{equation}
    \mathcal{E}_\text{NR-SSH}(k)=\begin{pmatrix}
    0 & (t_1-\gamma)+t_2 \text{e}^{\text{i} \kappa} \\
    (t_1+\gamma)+t_2 \text{e}^{-\text{i} \kappa} & 0    \end{pmatrix}
\end{equation}
%
%
where $\kappa$ is the dimensionless momentum. The spectrum of this Hamiltonian is
%
%
\begin{equation}
    \mathcal{E}_{\pm\text{NR-SSH}}(k)=\pm\sqrt{(t_1^2-\gamma^2)+t_2^2+2t_1t_2\cos\kappa+2\text{i}t_2\gamma\sin\kappa}.
\end{equation}
%
%
As in the case of the Hatano-Nelson model, it winds in the complex plane, see Figs.~\ref{fig_ICSE_SSH_Topo}\R{a} to~\ref{fig_ICSE_SSH_Topo}\R{d} and~\ref{fig_ICSE_SSH_Trivial}\R{a} to~~\ref{fig_ICSE_SSH_Trivial}\R{d}, thus we are expecting to observe NHSE. This is confirmed in Figs.~\ref{fig_ICSE_SSH_Topo}\R{e}, ~\ref{fig_ICSE_SSH_Topo}\R{f} and~\ref{fig_ICSE_SSH_Trivial}\R{e}, ~\ref{fig_ICSE_SSH_Trivial}\R{f}. For both the topological regime in Fig.~\ref{fig_ICSE_SSH_Topo}\R{e} and the trivial regime in Fig.~\ref{fig_ICSE_SSH_Trivial}\R{e}, we have fine-tuned the value of the impurity strength to almost cancel the NHSE on it. We note that there is a phase difference between the "A" and "B" sublattice, which is typical of the SSH model.
%
%
%
%
%
%
%
%


\comment{ection{non-Hermitian quantum walk}

In our study, we implement a discrete-time quantum walk on a one-dimensional lattice. The Hilbert space is given by
%
%
\begin{equation}
\mathcal{H} = \mathcal{H}_P \otimes \mathcal{H}_C,
\end{equation}
%
%
where $\mathcal{H}_P$ is the position space and $\mathcal{H}_C$ is the two-dimensional coin (or internal) space.

A single step of the quantum walk is defined by the unitary (or, in our case, non-unitary) evolution operator
%
%
\begin{equation}
U = S \, \left( I \otimes C \right),
\end{equation}
%
%
where $C$ is the coin operator acting on $\mathcal{H}_C$, and
$S$ is the conditional shift operator acting on the combined space.

The coin operator is chosen as
%
%
\begin{equation}
C = \begin{pmatrix}
\sqrt{r} & \sqrt{1-r} \\
\sqrt{1-\ell} & -\sqrt{\ell}
\end{pmatrix},
\end{equation}
%
%
with parameters $r,\ell \in [0,1]$ that determine the relative amplitudes for moving right or left, respectively. For the case $r=\ell$, the coin operator is unitary, and the walk is Hermitian. However, when $r\neq\ell$, the operator $C$ becomes non-unitary, thereby introducing non-Hermitian effects essential to mimic the asymmetric (non-reciprocal) hopping of the Hatano-Nelson model.

The shift operator is defined as
%
%
\begin{equation}
S = \sum_{x} \Big( |x+1\rangle\langle x| \otimes |R\rangle\langle R| + |x-1\rangle\langle x| \otimes |L\rangle\langle L| \Big),
\end{equation}
%
%
where $|R\rangle$ and $|L\rangle$ denote the coin states corresponding to right and left moves, respectively.

Thus, the state of the walker after $t$ steps is given by
%
%
\begin{equation}
|\psi(t)\rangle = U^t |\psi(0)\rangle.
\end{equation}
%
%
Due to the non-unitarity of $C$ for $r\neq\ell$, the overall evolution does not preserve norm, which effectively models gain or loss mechanisms in the system.

This non-Hermitian quantum walk provides a simple yet powerful framework to study phenomena such as asymmetric transport, the non-Hermitian skin effect, and impurity-induced localization. It thereby serves as a discrete model for the Hatano-Nelson system with an impurity.
In the following, we show how the dynamics of the quantum walk are modified by introducing the three different models of defects presented in Eqs.~(\ref{modA}--\ref{modC}) of the main text. In Fig.~\ref{QW_comparison}, we compare the quantum walk for an unbiased coin $r=\ell=1/2$ and for a non-Hermitian case $r>\ell$. We observe that in the unbiased case, the probability is symmetric with respect to the center of the 1D chain, whereas, for the non-Hermitian case, the likelihood of finding the walker on the right side of the chain is higher at the end of the evolution time. We note on passing that setting $r=\ell$ is equivalent to choose an unbiased coin.
%
%
    \begin{figure}
    \centering
    \begin{minipage}{0.45\textwidth}
        \includegraphics[width=\linewidth]{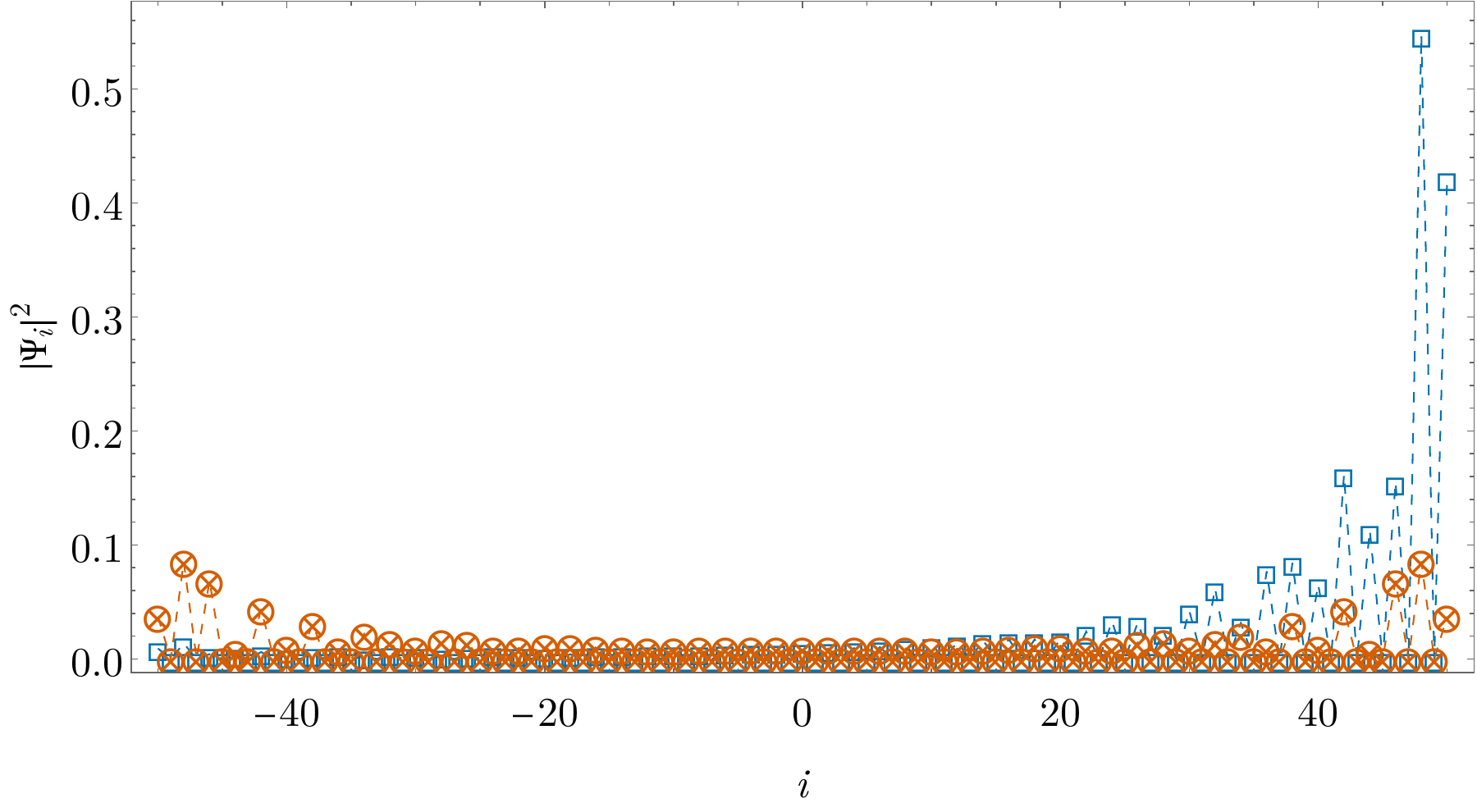}
    \end{minipage}
    \begin{minipage}{0.45\textwidth}
           \caption{\textbf{Comparison of quantum walks.} Comparison between two different Coins of quantum walks in the homogeneous case: The orange curve is for $r=\ell=1/2$, whereas the blue one is for $r=5.4$ and $\ell=0.5$. In both cases, we considered a linear chain with a length of 80 sites and 70 steps of evolution.}
            \label{QW_comparison}     
    \end{minipage}
\end{figure}    

%
%
In Fig.~\ref{QW_comparison_modABC}, we show the effects of the three model of impurity we have described in the main text. In the Fig.~\ref{QW_comparison_modABC}\R{a} and~\ref{QW_comparison_modABC}\R{d}, we consider the case of $\rm M_1$, whereas in Fig.~\ref{QW_comparison_modABC}\R{b} and~\ref{QW_comparison_modABC}\R{e} we consider $\rm M_2$, finally, in Fig.~\ref{QW_comparison_modABC}\R{c} and~\ref{QW_comparison_modABC}\R{f}, we consider $\rm M_3$. We clearly see that $\rm M_3$, is the best to \emph{block} the walker along the traveling to the side of the path favored by the non-Hermitian coin. We note that the value of the value of the impurity strength for $\rm M_2$ can be very detrimental if chosen equal to $\pi$, indeed for this specific case, it acts as an attractive potential trapping the walker at its position.
%
%
\begin{figure}[!b]
    \centering
    \includegraphics[width=0.85\textwidth]{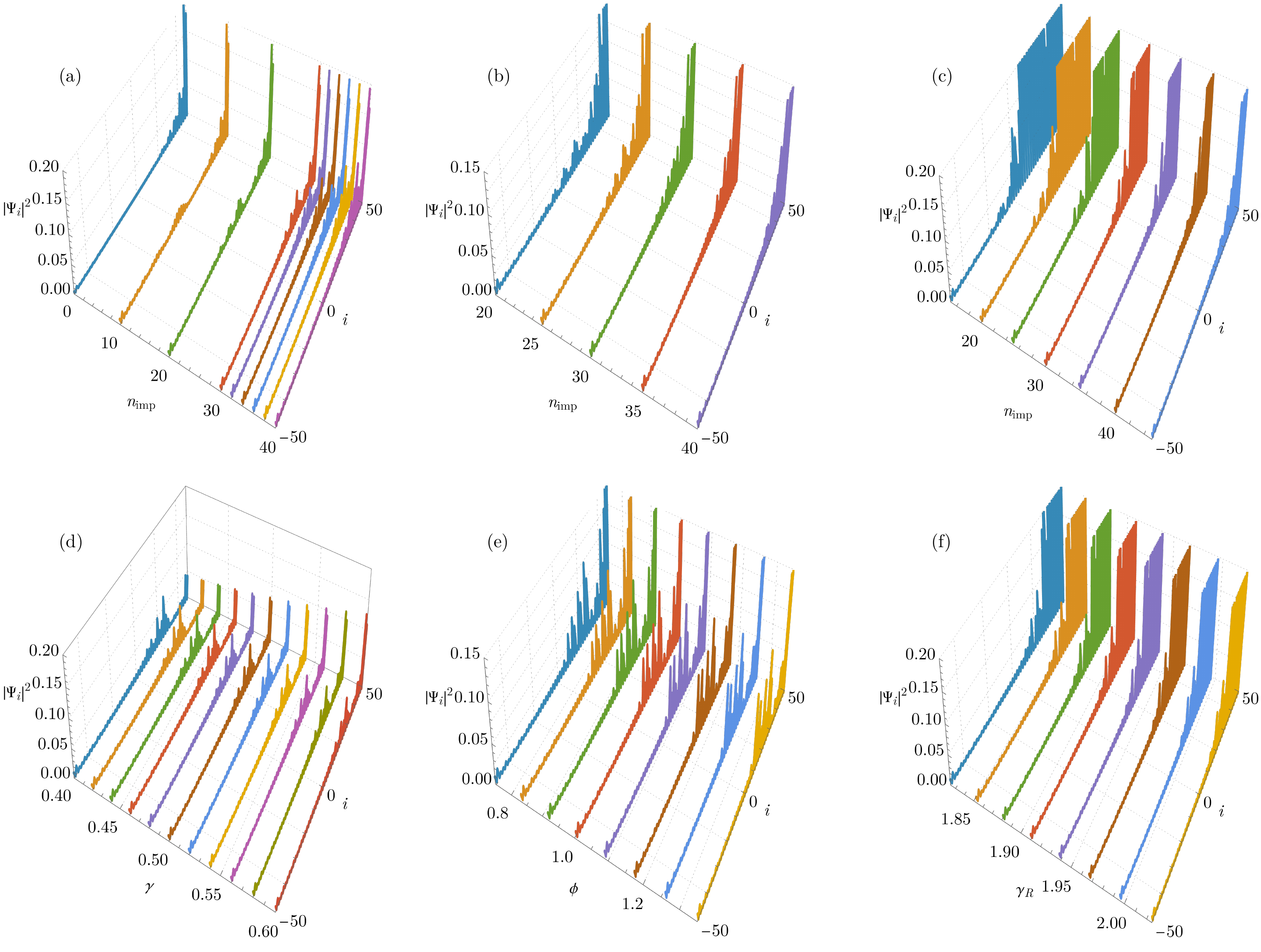}
    \caption{\textbf{Quantum walks with the defects, model comparison.} Comparison of the three possible implementations of onside defects for the non-Hermitian quantum walk. (a) $\rm M_1$ with $\gamma=0.7$ and various positions of the impurity $n_\text{imp}$. (b) $\rm M_2$ with $\phi=\pi/3$ and various positions of the impurity $n_\text{imp}$. (c) $\rm M_3$ with $\gamma_\text{L}=\gamma_\text{R}=(r\ell)^{-1/2}$ and various positions of the impurity $n_\text{imp}$. (d) $\rm M_1$ with $n_\text{imp}=30$ and various impurity strengths $\gamma$. (e) $\rm M_2$ with $n_\text{imp}=30$ and various impurity strengths $\phi$. (f) $\rm M_3$ with $n_\text{imp}=30$, $\gamma_\text{L}=(r\ell)^{-1/2}$ and various impurity strengths $\gamma_\text{R}$. In all the cases, we considered a linear chain with a length of 80 sites and 70 steps of evolution.}
    \label{QW_comparison_modABC}
\end{figure}
%
%
}

\end{document}